\pdfoutput=1
\documentclass[11pt,twoside,a4paper,cmspaper,final,collab]{cms-tdr}
\def\svnVersion{e3cb159}\def\svnDate{2024/10/18}\def\cmsCernNoTag{CERN-EP-2024-078}\def\cmsCernDate{\today}\def\cmsMessage{Published in Computing and Software for Big Science as \href{http://dx.doi.org/10.1007/s41781-024-00121-4}{\doi{10.1007/s41781-024-00121-4}.}}
\begin{document}\cmsNoteHeader{CAT-23-001}

\newlength\cmsFigWidth
\ifthenelse{\boolean{cms@external}}{\setlength\cmsFigWidth{0.99\columnwidth}}{\setlength\cmsFigWidth{0.65\textwidth}}
\ifthenelse{\boolean{cms@external}}{\providecommand{\cmsLeft}{upper\xspace}}{\providecommand{\cmsLeft}{left\xspace}}
\ifthenelse{\boolean{cms@external}}{\providecommand{\cmsRight}{lower\xspace}}{\providecommand{\cmsRight}{right\xspace}}

\newlength\cmsTabSkip\setlength{\cmsTabSkip}{3ex}

\providecommand{\cmsTable}[1]{\resizebox{\textwidth}{!}{#1}}
\newcommand{\ttseq}[1]{\texttt{\seqsplit{{#1}}}}
\newcommand{\combine}{\textsc{Combine}\xspace}
\newcommand{\Combine}{\textsc{Combine}\xspace}
\newcommand{\altcombine}{{\normalfont\scshape Combine}\xspace}
\newcommand{\Root}{\textsc{ROOT}\xspace}
\newcommand{\ttree}{\ttseq{TTree}\xspace}
\newcommand{\RooFit}{\textsc{RooFit}\xspace}
\newcommand{\Roofit}{\textsc{RooFit}\xspace}
\newcommand{\RooStats}{\textsc{RooStats}\xspace}
\newcommand{\cmssw}{\textsc{cmssw}\xspace}
\newcommand{\vtheta}{\ensuremath{\vec{\nu}}\xspace}
\newcommand{\Likelihood}{\ensuremath{\mathcal{L}}\xspace}
\newcommand{\ttheta}{\ensuremath{y}\xspace}
\newcommand{\yield}{\ensuremath{\omega}\xspace}
\newcommand{\anuisance}{\ensuremath{\nu}\xspace}
\newcommand{\knuisance}{\ensuremath{\nu_{k}}\xspace}
\newcommand{\aknuisance}{\ensuremath{\nu_{k}}\xspace}
\newcommand{\tthetak}{\ensuremath{y_{k}}\xspace}
\newcommand{\tvtheta}{\ensuremath{\vec{y}}\xspace}
\newcommand{\qxobsm}{\ensuremath{q_{\text{x}}^{\text{obs}}(\mu)}\xspace}
\newcommand{\rqqH}{\ensuremath{r_{\PQq\PQq\PH}}\xspace}
\newcommand{\rggH}{\ensuremath{r_{\Pg\Pg\PH}}\xspace}

\definecolor{codegreen}{rgb}{0,0.6,0}
\definecolor{codegray}{rgb}{0.5,0.5,0.5}
\definecolor{codepurple}{rgb}{0.58,0,0.82}
\definecolor{backcolour}{rgb}{.97,.97,.97}
\definecolor{datacardcolor}{rgb}{1.,.98,.80}
\definecolor{codeoutputcolor}{rgb}{0.729,0.969,0.745}
\definecolor{commandlinecolor}{rgb}{0.847,0.749,0.847}

\lstdefinestyle{mystyle}{
    backgroundcolor=\color{backcolour},
    keywordstyle=\color{orange},
    numberstyle=\tiny\color{codegray},
    stringstyle=\color{codepurple},
    basicstyle=\ttfamily\bfseries\footnotesize,
    breakatwhitespace=false,
    breaklines=true,
    frame=single,
    keepspaces=true,
    showspaces=false,
    showstringspaces=false,
    showtabs=false,
    tabsize=2,
    breakautoindent=true,
    breakindent=2ex,
    postbreak=\mbox{\textcolor{black}{$\hookrightarrow$}\space}
}

\lstdefinestyle{output}{
    backgroundcolor=\color{codeoutputcolor},   
    keywordstyle=\color{orange},
    numberstyle=\tiny\color{codegray},
    stringstyle=\color{codepurple},
    basicstyle=\ttfamily\bfseries\footnotesize,
    breakatwhitespace=false,         
    breaklines=true, 
    frame=single,
    keepspaces=true,                 
    showspaces=false,                
    showstringspaces=false,
    showtabs=false,                  
    tabsize=2,
    breakautoindent=true,
    breakindent=2ex,
    postbreak=\mbox{\textcolor{black}{$\hookrightarrow$}\space}
}
\lstdefinestyle{commandline}{
    backgroundcolor=\color{commandlinecolor},   
    keywordstyle=\color{orange},
    numberstyle=\tiny\color{codegray},
    stringstyle=\color{codepurple},
    basicstyle=\ttfamily\bfseries\footnotesize,
    breakatwhitespace=false,         
    breaklines=true, 
    frame=single,
    keepspaces=true,                 
    showspaces=false,                
    showstringspaces=false,
    showtabs=false,                  
    tabsize=2,
    breakautoindent=true,
    breakindent=2ex,
    postbreak=\mbox{\textcolor{black}{$\hookrightarrow$}\space}
}

\lstdefinestyle{datacard}{
    backgroundcolor=\color{datacardcolor},   
    keywordstyle=\color{orange},
    numberstyle=\tiny\color{codegray},
    stringstyle=\color{codepurple},
    basicstyle=\ttfamily\bfseries\small,
    breakatwhitespace=false,
    breaklines=true,
    frame=single,
    keepspaces=true,
    numbers=left,
    showspaces=false,
    showstringspaces=false,
    framexleftmargin=2em,
    xleftmargin=2em,
    showtabs=false,
    tabsize=2,
    linewidth=\textwidth,
    breakautoindent=true,
    breakindent=2ex,
    postbreak=\mbox{\textcolor{black}{$\hookrightarrow$}\space}
}

\lstdefinestyle{datacardinline}{
    backgroundcolor=\color{datacardcolor},   
    keywordstyle=\color{orange},
    numberstyle=\tiny\color{codegray},
    stringstyle=\color{codepurple},
    basicstyle=\ttfamily\bfseries\footnotesize,
    breakatwhitespace=false,
    breaklines=true,
    frame=single,
    keepspaces=true,
    numbers=left,
    showspaces=false,
    showstringspaces=false,
    framexleftmargin=2em,
    xleftmargin=2em,
    showtabs=false,
    tabsize=2,
    postbreak=\mbox{\textcolor{black}{$\hookrightarrow$}\space}
}

\renewcommand\lstlistingname{Datacard}
\renewcommand\lstlistlistingname{Datacard}

\hyphenation{Data-card}
\hyphenation{data-card}
\hyphenation{data-set}
\hyphenation{param-eter}
\hyphenation{Roo-Real-Var}
\hyphenation{al-pha}
\hyphenation{Roo-Workspace}
\hyphenation{Roo-Abs-Real}
\hyphenation{Roo-Abs-Pdf}
\hyphenation{Roo-Data-Set}
\hyphenation{Roo-Data-Hist}

\cmsNoteHeader{CAT-23-001}
\title{The CMS statistical analysis and combination tool: \altcombine}
\date{\today}

\abstract{
This paper describes the \combine software package used for statistical analyses by the CMS Collaboration. The package, originally designed to perform searches for a Higgs boson and the combined analysis of those searches, has evolved to become the statistical analysis tool presently used in the majority of measurements and searches performed by the CMS Collaboration. It is not specific to the CMS experiment, and this paper is intended to serve as a reference for users outside of the CMS Collaboration, providing an outline of the most salient features and capabilities. Readers are provided with the possibility to run \combine and reproduce examples provided in this paper using a publicly available container image. Since the package is constantly evolving to meet the demands of ever-increasing data sets and analysis sophistication, this paper cannot cover all details of \combine. However, the online documentation referenced within this paper provides an up-to-date and complete user guide. 
}
\hypersetup{
pdfauthor={CMS Collaboration},
pdftitle={The CMS statistical analysis and combination tool: Combine},
pdfsubject={CMS},
pdfkeywords={CMS, combine, Higgs, Higgs boson, statistics}
}

\provideboolean{cms@use-overleaf}
\provideboolean{cms@pas}
\ifthenelse{\boolean{cms@external}}{\providecommand{\cmsLeft}{upper\xspace}}{\providecommand{\cmsLeft}{left\xspace}}
\ifthenelse{\boolean{cms@external}}{\providecommand{\cmsRight}{lower\xspace}}{\providecommand{\cmsRight}{right\xspace}}

\author{The CMS Collaboration}

\maketitle

\clearpage

\section{Introduction}
\label{sec:introduction}
The CMS statistical analysis software, \combine, is designed with two main features in mind. 
The first is to provide a command-line interface to several common workflows used in statistical analyses in high-energy physics, and the second is to encapsulate the statistical model using a human-readable configuration file -- herein referred to as a ``datacard''.
These features are intended to ensure consistency in statistical methodology and allow for efficient investigation of potential issues, without limiting the complexity of any single analysis.
Perhaps the most important consequence is that the constructed likelihoods can be combined to produce a greater sensitivity in searches or measurements, provided that the data sets are statistically independent.
The \combine analysis software is built around the \Root~\cite{Brun:1997pa}, \RooFit~\cite{Verkerke:2003ir}, and \RooStats~\cite{Verkerke:2003ir} packages.

The statistical methods in \combine, many of which were developed in the LHC Higgs Combination Group~\cite{LHC-HCG}, were originally designed for searches for a Higgs boson in proton-proton collisions. 
The \combine tool was used in early searches~\cite{CMS:2012zhx} for, and the subsequent discovery of, the Higgs boson~\cite{Chatrchyan:2012ufa,Chatrchyan:2013lba} by the CMS Collaboration. 
Historically, its main use case within the CMS Collaboration was in searches for the Higgs boson, hence many of the function names and variables used in \combine include ``Higgs''.
However, the tool is not specific to these searches or measurements of Higgs boson properties but instead is a generic tool usable for various statistical analyses of LHC data.
Since the Higgs boson discovery, many extensions have been included that have been used for the statistical analysis in numerous publications of the CMS Collaboration, including measurements of Higgs boson properties~\cite{Sirunyan:2018koj}, searches for supersymmetry~\cite{CMS:2021eha}, and measurements of standard model parameters such as the top quark mass~\cite{CMS:2022emx}. 
The \combine tool has also been used with data from the ATLAS and CMS experiments to produce combined measurements of the Higgs boson mass, production and decay rates, and coupling modifiers~\cite{ATLAS:2015yey,ATLAS:2016neq}. 
Furthermore, the \combine software includes several routines that provide diagnostic information regarding the statistical model and statistical analysis methods used in such publications.  

This paper provides a summary of the statistical methods and capabilities of the \combine tool.
For complete and up-to-date documentation, it is recommended that the reader consult the online documentation~\cite{combinedoc}.
 
In this paper, command line instructions are indicated by the symbol \texttt{\$} at the start of the line.
Square brackets \texttt{[option]} indicate optional commands that alter the default behavior of \combine, while angular brackets  \ttseq{<option value>} indicate a value that must be specified by the user. 
The color scheme for the listings contained in this paper is as follows:
\begin{lstlisting}[style=datacardinline]
Contents of a complete datacard.
\end{lstlisting}
\begin{lstlisting}[style=commandline]
$ Executable command line.
\end{lstlisting}
\begin{lstlisting}[style=output]
> Terminal output from the tool.
\end{lstlisting}
\begin{lstlisting}[style=mystyle]
Snippet of a datacard or of Python code.
\end{lstlisting}

This paper is organized as follows. 
Section~\ref{sec:installation} details the dependencies of the software and instructions for its installation. 
The statistical model that is constructed by \combine is described in 
Section~\ref{sec:thestatisticalmodel}, followed by detailed explanations of the analysis types available in the tool and instructions on how they are implemented in Section~\ref{sec:datacard}. 
Section~\ref{sec:phys_model} provides instructions on the use of physics models in \combine, with several examples given.
This section also provides a concrete example of a full statistical model constructed in \combine.
Section~\ref{sec:runningthetool} provides instructions on how to run \combine. 
It demonstrates several common statistical procedures using the tool, for example, the calculation of maximum likelihood estimates and confidence or credible intervals, and performing goodness of fit tests.
Finally, a summary is given in Section~\ref{sec:summary}.

\section{Installation}\label{sec:installation}
Aside from \Root and \RooFit, and their dependencies, several additional libraries are used for optimized algebraic calculations, such as vectorization libraries \textsc{vdt}~\cite{Piparo_2014}, the GNU scientific library \textsc{gsl}~\cite{galassi2018}, and \textsc{eigen}~\cite{eigen}. 
Additional common libraries are used, which include \textsc{boost}~\cite{boost}, and \textsc{gzip}~\cite{gzip}.
The \combine package may be compiled either within a CMS software (\cmssw) environment that provides a versioned set of all dependencies, or as a standalone package.
Details of different installation instructions are regularly updated in the online documentation.
A precompiled version of \combine is available as a \textsc{Docker}~\cite{merkel2014docker} container image: 
\begin{lstlisting}[style=commandline]
$ docker run [--platform linux/amd64] --name combine -it gitlab-registry.cern.ch/cms-cloud/combine-standalone:v9.2.0
\end{lstlisting}
At the time of writing this paper, the latest version of \combine is v9.2.0, and all of the example datacards and the inputs necessary to run \combine can be found in the \ttseq{data/tutorials/CAT23001} directory, which is available in the container image.  
For statistical calculations that make use of random sampling, the results obtained by a reader are expected to be consistent with, but not identical to, those provided in this paper.  

\section{The statistical model}\label{sec:thestatisticalmodel}
The primary task of \combine is to produce a statistical model, $p(\text{data};\vec{\Phi})$, which encodes the probability density for the observed data parameterized by the model parameters $\vec{\Phi}$, and is subsequently used for the statistical analysis.

For numerical efficiency, it is useful to factorize the statistical model $p(\text{data};\vec{\Phi})$ as much as possible with respect to both observables and parameters. 
The parameter space is partitioned into parameters of interest $\vec{\mu}$ and nuisance parameters $\vtheta$.
Nuisance parameters are used to model various uncertainties of theoretical and experimental origin, such as those involved in the prediction of process cross sections or associated with luminosity calibration.
Furthermore, the observable space is partitioned into primary observables $\vec{x}$, defined as those that appear in components of the model containing the parameters of interest, and auxiliary observables $\tvtheta$ that appear only in components of the model containing nuisance parameters. 
Each nuisance parameter $\knuisance$ (each element of the vector $\vtheta$) is paired with a corresponding auxiliary observable $\ttheta_{k}$ (element of $\tvtheta$) in a component of the statistical model that provides information about how well the nuisance parameter is known. 
Therefore, the statistical model constructed using \combine is factorized into the primary and auxiliary components of the probability as 
\begin{linenomath*}
\begin{equation}\label{eqn:stat-model}
  p({\vec{x},\tvtheta;\vec{\Phi}}) = p(\vec{x};\vec{\mu},\vtheta) \prod_{k}p_{k}(\tthetak;\knuisance),
\end{equation}
\end{linenomath*}
where $p(\vec{x};\vec{\mu},\vtheta)$ is the probability distribution of the observables for the primary analysis, and $p_{k}(\tthetak;\knuisance)$ are the probability distributions of the auxiliary observables. 
A likelihood function is constructed for a particular data set of independent identically distributed observables $\left\{\vec{x}_{d}\right\}$ as
\begin{linenomath*}
  \begin{equation}\label{eqn:lh-model}
    \Likelihood(\vec{\Phi}) =   \prod_{d}p(\vec{x}_{d};\vec{\mu},\vtheta)\prod_{k}p_{k}(\tthetak;\knuisance),
\end{equation}
\end{linenomath*}
where the subscript $d$ runs over all entries in the data set and the likelihood function is used in both Bayesian and frequentist calculations. 
Early searches for the Higgs boson by the CMS collaboration reported upper limits on the Higgs boson cross section using both Bayesian and frequentist methods~\cite{CMS:2012zhx} with \combine.  

In \combine, the probability density terms associated with the auxiliary observables, $p_{k}(\tthetak;\knuisance)$, can also be reinterpreted as posterior distributions for the nuisance parameters, $p_{k}(\knuisance|\tthetak)$, resulting from the outcome of measurements of, or otherwise justified constraints on, the auxiliary observables $\tthetak$, through the relationship
\begin{linenomath*}
\begin{equation}\label{eq:priors}
  p_{k}(\knuisance|\tthetak) \propto p_{k}(\tthetak;\knuisance)\pi_{k}(\knuisance),
\end{equation} 
\end{linenomath*}
where $\pi_{k}(\knuisance)$ are the nuisance parameter priors. 
This procedure provides probability distributions from which nuisance parameter values can be sampled when generating pseudo-data sets that \combine uses in certain statistical calculations, as described in Section~\ref{toy-data-generation}. 
For all types of nuisance parameters in \combine, the priors are always assumed to be uniform~\cite{LHC-HCG} so all of the $\pi_{k}(\knuisance)$ are constants. 

Each element of $\vec{x}$ is referred to as a ``channel'' and is statistically independent from all other elements of $\vec{x}$. 
For example, each element of $\vec{x}$ could be the event counts in different reconstructed final states of some data set, or a continuous observable such as the invariant mass of a pair of final state particles. 
The term $p(\vec{x};\vec{\mu},\vtheta)$ in Eq.~(\ref{eqn:stat-model}) becomes a product over the channels,
\begin{linenomath*}
  \begin{equation}\label{eqn:channel-model}
    p(\vec{x};\vec{\mu},\vtheta) = \prod_{i} p_{i}({x}_{i};\vec{\mu},\vtheta),
\end{equation}
\end{linenomath*}
where $i$ runs over the channels that comprise the primary analysis and $p_{i}$ is the probability density function (pdf) for the observable ${x}_{i}$.

The likelihood function constructed by \combine assumes that all $\tthetak$ are statistically independent from each other and from the primary observables.
The user must specify the observables and their pdfs using the datacard as described below. 

\section{Supported analysis types}
\label{sec:datacard}
A configuration file in plain text format is required for \combine to define the observables $\vec{x}$ and $\tvtheta$, and their pdfs $p(\vec{x};\vec{\mu},\vtheta)$ and $p_{k}(\tthetak;\knuisance)$.
This file is the datacard and is the primary input to \combine.
The file is also used to specify the observed data needed to define the likelihood function, whether the analysis is a simple counting experiment or a more complex analysis using binned or unbinned distributions of the data with histograms or parametric functions to describe the pdfs.

The package includes a script \ttseq{text2workspace.py} that \combine uses to convert the user defined inputs (in the form of the datacard) into a binary representation of the statistical model.
The script can be run before running \combine itself to produce a binary \Root file containing the statistical model in the form of a \RooFit \ttseq{RooWorkspace} object. 
The script is  automatically run if the datacard is provided as the input to \combine. 
The script is run with the following command:
\begin{lstlisting}[style=commandline]
$ text2workspace.py [-m <mass>] [-o <datacard>.root] <datacard>.txt
\end{lstlisting}
The  output \Root file is given the same name as the input datacard, with the extension modified to \texttt{.root} unless the option \texttt{-o} is specified.
The value of \texttt{mass}, a parameter widely used in searches for new particles, is interpreted by the \ttseq{text2workspace.py} script to specify the datacard keyword \ttseq{\$MASS}, as described in Section~\ref{sec:template-based}.

It is possible to combine several datacards into a single datacard using the \ttseq{combineCards.py} script:
\begin{lstlisting}[style=commandline]
$ combineCards.py Name1=card1.txt Name2=card2.txt .... > <combined card>.txt
\end{lstlisting}
This allows for building complex statistical models, while retaining the readability of individual components (datacards) of the model.
Multiple instances of any nuisance parameter, sharing the same name, are treated as a single parameter of the statistical model with a single corresponding auxiliary observable $\ttheta$, provided that the pdf specified for $\ttheta$ is the same in each instance.   
The rest of this section describes the preparation of datacards and associated inputs for use with \combine.

The first line of the datacard is a declaration of the number of channels, \texttt{imax}, that are present in the statistical model: 
\begin{lstlisting}[style=mystyle]
imax <number of channels>
\end{lstlisting}

For a single-channel analysis the datacard entry would be \texttt{imax 1}. 
If the value of \texttt{imax} is specified as ``\texttt{*}'', \combine automatically determines the number of channels.

The next lines in the datacard declare the number of processes to be considered, \texttt{jmax}$+1$, and the number of nuisance parameters, \texttt{kmax}:
\begin{lstlisting}[style=mystyle]
jmax <number of processes minus one>
kmax <number of nuisance parameters or sources of systematic uncertainties>
\end{lstlisting}
For datacards with a single signal process, \texttt{jmax} is the number of background processes. 
Datacard lines starting with ``\texttt{\#}'' are ignored by \combine and any amount of whitespace is allowed to separate columns and lines in the datacard.
These features can be used to include descriptive comments in the datacard.
The next sections of the datacard have a different syntax depending on whether the datacard represents a counting or shape analysis. 

\subsection{Counting analyses}
A counting analysis is one for which the statistical model can be cast in the form of Eq.~(\ref{eqn:stat-model}) with only one primary observable, namely the total event count in a single channel that includes multiple sources of signal and background. 
In the following, the primary observable is labeled $n$.
The probability to observe $n$ events is described by a Poisson distribution, 
\begin{linenomath*}
  \begin{equation}\label{eqn:lh-model2}
    \mathcal{P}(n;\lambda) =  \lambda^{n}\frac{\re^{-\lambda}}{n!},
\end{equation}
\end{linenomath*}
for which the expected value, $\lambda$, can be a function of one or more parameters, and represents the total number of expected signal and background events.

Each process comes with a specified reference rate and one or more sources of uncertainty that are referred to as ``systematic'', even if they are of statistical origin. 
In \combine the definitions of each systematic uncertainty  require  the functional form for the probability distribution $p_{k}(\tthetak;\knuisance)$ to be specified. 
These typically reflect calibration measurements that often result in log-normal or gamma distributions, and are implemented in \combine through multiplicative factors that propagate the effect of systematic uncertainties to the statistical model. 

Datacard~\ref{dc:counting} is an example with all of these elements, representing a counting experiment with one channel having a signal process ${\Pp\Pp\mathrm{X}}$ with reference rate of 1.47, two background processes $\PW\PW$ and $\PQt\PAQt$, and three nuisance parameters that model systematic uncertainties in both the signal and background rates.  
In this example, the integrated luminosity uncertainty (\ttseq{lumi}), assumed to be log-normal, results in an uncertainty in the expected signal rate, as well as in the expected rates of the $\PW\PW$ and $\PQt\PAQt$ backgrounds. 
The reference rates of the signal and background processes are determined using simulation. 
A log-normal type uncertainty in the signal rate (\texttt{xs}) is included to account for the uncertainty in the predicted cross section of the signal process. 
A limited number of simulated events are available to determine the rate of the $\PW\PW$ background. 
The statistical uncertainty due to this limited number of simulated events (\texttt{nWW}) propagates as a gamma distribution to the rate of $\PW\PW$. 

The datacard lines immediately following the \texttt{imax}, \texttt{jmax}, and \texttt{kmax} lines describe the number of events observed in each channel. 
Line number 5, starting with \texttt{bin}, defines the label that should be used for each channel. 
In this example there is one channel, labeled \texttt{ch1}. 
Line number 6, starting with the word \ttseq{observation}, indicates the number of observed events, which is 0 in Datacard~\ref{dc:counting}. 
For analyses in which the data are binned in a histogram, a template-based datacard can be used instead of treating each bin of the histogram as a separate channel.
\begin{figure*}[t]  
\begin{lstlisting}[style=datacard,label=dc:counting,caption=Counting experiment datacard - \texttt{datacard-1-counting-experiment.txt}\vspace{0.25\baselineskip}]
imax 1
jmax 2
kmax 3
# A single channel - ch1 - in which 0 events are observed in data
bin            ch1
observation    0
# ----------
bin            ch1   ch1   ch1
process        ppX   WW    tt
process        0     1     2     
rate           1.47  0.64  0.22
# ----------
lumi    lnN    1.11  1.11   1.11
xs      lnN    1.20    -     - 
nWW     gmN 4    -   0.16    - 
\end{lstlisting}
\end{figure*} 
There are typically several processes that contribute to the overall signal or background expected yields.
Lines 8--11 in Datacard~\ref{dc:counting} describe the number of events expected for each channel and process, arranged in columns. 
The first column in each row identifies the information expected in the remaining columns. 
The number of columns beyond the first column must be equal to the total number of processes across all channels, i.e, to the product $\texttt{imax} (\texttt{jmax}+1)$. 
Line 8 starting with \texttt{bin} indicates that this row specifies the channel that each column refers to. 
In this case, since there is only a single channel, the number of columns in addition to the first one is equal to the number of signal and background processes in this channel. 
Lines 9 and 10 starting with \texttt{process} indicate that these rows refer to the labels and types of the various processes. 
Line 9 provides a label for each process and line 10 defines the type of the process which is either a positive number for  a background process, or \texttt{0} or a negative number for a signal process.
Line 11, starting with \texttt{rate}, indicates the expected event yield in the specified channel and process. 
This value should be considered as a reference rate for the process, assuming predetermined values for theoretical cross sections, detector acceptance and selection efficiencies, and integrated luminosity of the data set used in the analysis. 
The rates can be modified by the parameters of the statistical model $\vec{\Phi}$.  
In the simplest statistical model available in \combine, a single parameter of interest, the signal strength $r$, multiplies the rate of every signal process in the datacard as described in Section~\ref{sec:phys_model}.

The remaining lines 13--15 contain the description of systematic uncertainties that are to be included in the statistical model.
Each of these systematic uncertainties is associated with a dedicated nuisance parameter $\anuisance$.
The systematic uncertainties section of the datacard is structured as follows:
\begin{itemize}
    \item   The first column indicates the name used in the binary representation of the statistical model for identifying the uncertainty. This is the name given to the \RooFit \ttseq{RooRealVar} object that encodes the corresponding nuisance parameter in the statistical model.
    \item   The second column identifies the effect of the associated nuisance parameter and the form of $p(\ttheta;\anuisance)$ to be included in the statistical model. For gamma type nuisance parameters, this column has two entries, which is explained in the following. 
    \item   Finally, there are columns describing the effect of the systematic uncertainty on the rate of each process in each channel. The number of columns is the same as for the previous lines declaring channels, processes, and rates. If a process is unaffected by a nuisance parameter, the corresponding column entry for that nuisance parameter is ``\texttt{-}''. 
\end{itemize}
The different types of systematic uncertainties that can be included in the datacard for counting experiments are shown in Table~\ref{tab:countingsys}.
Each of these types results in an associated probability term $p(\ttheta;\anuisance)$ which is either a normal distribution,
\begin{linenomath*}
\begin{equation}\label{eqn:normaldist}
  \mathcal{N}(\ttheta;\anuisance,\sigma_{\anuisance}) = \frac{1}{\sigma_{\anuisance}\sqrt{2\pi}}\re^{-\frac{1}{2}\left(\frac{\anuisance-\ttheta}{\sigma_{\anuisance}}\right)^{2}},
\end{equation}
\end{linenomath*}
Poisson distribution,
\begin{linenomath*}
\begin{equation}\label{eqn:poissondist}
\mathcal{P}(\ttheta;\anuisance) = \anuisance^{\ttheta}\frac{\re^{-\anuisance}}{\ttheta!},
\end{equation}
\end{linenomath*}
or uniform distribution,
\begin{linenomath*}
\begin{equation}
  \mathcal{U}(\ttheta;a,b) = \begin{cases}
    \frac{1}{b-a} & \text{if } \ttheta\in [a,b],\\
    0 & \text{otherwise}. 
  \end{cases}
\end{equation}
\end{linenomath*}
Equation~(\ref{eqn:normaldist}), aside from being used directly for normally distributed observables, is also the building block for log-normally distributed observables, as discussed below. 
The Poisson distribution of $\ttheta$ in Eq.~(\ref{eqn:poissondist}) becomes a gamma distribution $\Gamma(\anuisance;\ttheta)$ when the observed value of $\ttheta$ is substituted and the expression is interpreted as a likelihood function, 
\begin{linenomath*}
\begin{equation}
\Gamma(\anuisance;\ttheta) = \anuisance^{\ttheta}\frac{\re^{-\anuisance}}{\ttheta!}.  
\end{equation}
\end{linenomath*}

In Datacard~\ref{dc:counting} there are three sources of systematic uncertainty that affect one or more processes.
Each of these lines results in a single nuisance parameter $\anuisance$, auxiliary observable $\ttheta$, and associated probability density $p(\ttheta;\anuisance)$ being included in the statistical model. 
The nuisance parameters are $\anuisance_{\texttt{lumi}},~\anuisance_{\texttt{xs}}$, and $\anuisance_{\texttt{nWW}}$ with corresponding auxiliary observables $\ttheta_{\texttt{lumi}},~\ttheta_{\texttt{xs}}$, and $\ttheta_{\texttt{nWW}}$.

The first two uncertainties are log-normal types~\cite{James:1019859}, detailed below using the integrated luminosity $L$ uncertainty as an example.
The rates of signal and background are typically proportional to $L$. 
The rates defined in the datacard are normalized to a reference value $L_0$, which represents the nominal value of the integrated luminosity. Deviations from this reference value are expressed through the dimensionless quantity $f = L/L_{0}$. 
An estimate $\hat{f}$ is available as a random sample from a pdf $p(\hat{f};f)$. 
The probability distribution $p(\hat{f};f)$ is log-normal, so that the sampling distribution of $\ln \hat{f}$ is normal, with mean equal to $\ln f$. 
The standard deviation of $\ln \hat{f}$ can be judiciously written as $\ln \kappa$, where $\kappa$ is a positive constant that is specified in the datacard.
The nuisance parameter $\anuisance$ that controls the systematic uncertainty in the integrated luminosity is then not considered to be $L$ itself, but rather is defined as $\anuisance = \ln f / \ln \kappa$ from which it follows that the multiplicative factor in the rate corresponding to luminosity is $f(\anuisance) = \kappa^{\anuisance}$. 
The observable $\ttheta$ is then a sample from $p(\ttheta;\anuisance)$, which is normal with mean $\anuisance$ as in Eq.~(\ref{eqn:normaldist}), and unity standard deviation.
The log-normal type is typically used when $f$ is positive by definition, as is the case with the integrated luminosity, and $p(\hat{f};f)$ continuously approaches zero as $\hat{f}\to 0$.

The first uncertainty in Datacard~\ref{dc:counting}, \texttt{lumi} in line 13, represents the uncertainty in the measured integrated luminosity of the data set. 
The effect of this uncertainty is 11\% on the rate of the \texttt{ppX}, \texttt{WW}, and \texttt{tt} processes.
This means that the rates of all three processes are multiplied by a factor of 1.11 when the nuisance parameter $\anuisance_{\texttt{lumi}}$ is set to $+1$, and by a factor of $1/1.11=0.90$ when it is set to $-1$. 
The log-normal type is also useful when a systematic uncertainty is taken to be a factor of $\kappa$, also often expressed as percentage uncertainty of $\kappa-1$. 
This means that high tails of the distribution of $\hat{f}$ above $\kappa f$ and low tails below $f/\kappa$ each contain equal probability of roughly 16\%. 
The second uncertainty in Datacard~\ref{dc:counting}, \ttseq{xs} in line 14, represents an uncertainty in the calculation of the theoretical cross section of the signal process. 
This uncertainty has an effect of 20\% on the signal rate while leaving the background processes unaffected. 

The gamma type uncertainty is used to model uncertainties in the rate of a particular process due to the limited sample size used to predict the rate. The third uncertainty in Datacard~\ref{dc:counting}, \ttseq{nWW} in line 15, is a gamma type uncertainty. 
This line in the datacard specifies that the reference rate of the \texttt{WW} background process is determined from a sample of 4 simulated events, each with an event weight of 0.16.

\begin{table*}[ht!]
    \topcaption{Available uncertainty types for counting experiments.
    The second and third columns indicate the entries for the datacard required to specify the type, and the relative effect on the yield of each process in each channel.
    The fourth and fifth columns indicate the resulting multiplicative factor by which \combine scales the normalization of the relevant process in the specified channel, and the term $p(\ttheta;\anuisance)$ that is included in Eq.~(\ref{eqn:stat-model}).
    Finally, the last column indicates the default values of $\anuisance$ and $\ttheta$.
    Where relevant, the value of $\kappa-1$ can be interpreted as the relative uncertainty in the process normalization in a given channel.
    }
    \newcommand{\mylittlecell}[3][]{\begin{tabular}[#1]{@{}#2@{}}#3\end{tabular}}
    \renewcommand{\thempfootnote}{\ifcase\value{mpfootnote}\or\ensuremath{\star}\or\ensuremath{\dagger}\or\ensuremath{\ddagger}\fi}
    \centering
    \renewcommand{\arraystretch}{1.6}
    \begin{minipage}{\textwidth}
    \cmsTable{
    \begin{tabular}{lccccc}
    {Uncertainty type} & {Directive}  & {Inputs} & {Multiplicative factor}, $f(\anuisance)$ & $p(\ttheta;\anuisance)$  & {Default values}
    \\
    \hline
    Log-normal & \texttt{lnN} & \texttt{kappa} & $\kappa^{\anuisance}$ & $\mathcal{N}(\ttheta;\anuisance,1)$ & $\anuisance=\ttheta=0$ \\ [\cmsTabSkip]
    \mylittlecell{l}{Asymmetric \\ log-normal}& \texttt{lnN} &   \mylittlecell{c}{\texttt{kappaDown},\\\texttt{kappaUp}} & \mylittlecell{c}{$\left(\kappa^{\text{Down}}\right)^{-\anuisance}$ if $\anuisance < -0.5$,\\ $\left(\kappa^{\text{Up}}\right)^{\anuisance}$ if $\anuisance > 0.5$, \\ $\re^{\anuisance K\left(\kappa^{\text{Down}},\kappa^{\text{Up}},\anuisance\right)}$ otherwise.\footnote{\scriptsize $K\left(\kappa^{\text{Down}},\kappa^{\text{Up}},\anuisance\right) = \frac{1}{8} \left[4\ln\left(\kappa^{\text{Up}}/\kappa^{\text{Down}}\right)+\ln\left(\kappa^{\text{Up}}\kappa^{\text{Down}}\right) \left(48\anuisance^5 - 40\anuisance^3 + 15\anuisance\right) \right]$ ensures that the multiplicative factor and its first and second derivatives are continuous for all values of $\anuisance$, and reduces to a log-normal for $\kappa^{\text{Down}}=1/\kappa^{\text{Up}}$.}} & $\mathcal{N}(\ttheta;\anuisance,1)$ & $\anuisance=\ttheta=0$ \\ [\cmsTabSkip]
    Log-uniform & \texttt{lnU} & \texttt{kappa} &
    $\kappa^{\anuisance}$
    & $\mathcal{U}\left(y,1/\kappa,\kappa\right)$ & $\anuisance=\ttheta=\frac{1}{2}\left(\kappa+1/\kappa\right)$ \\ [\cmsTabSkip]
    Gamma & \texttt{gmN} & \texttt{N}, \texttt{alpha}\footnote{\scriptsize The rate value for the affected process must be equal to $N\alpha$.} &
    $\anuisance/N$
    & $\mathcal{P}(\ttheta;\anuisance)$ & $\anuisance=N+1$, $\ttheta=N$\footnote{\scriptsize The default value for the nuisance parameter is set to the mean of a gamma distribution with parameters $\kappa=N+1,~\lambda=1$, as defined in Ref.~\cite{Workman:2022ynf}.}\\
    \end{tabular}
    }
    \end{minipage}
    \label{tab:countingsys}
\end{table*}

The full statistical model that \combine produces by default using Datacard~\ref{dc:counting} is given in Section~\ref{sec:phys_model}. 

\subsection{Shape analyses}
\label{sec:shapeana}
A shape analysis is defined as one that incorporates one or more primary observables, beyond a single number of events, in the statistical model of Eq.~(\ref{eqn:stat-model}). 
The datacard has to be supplemented with two extensions: a new block of lines defining the pdfs for the observables related to each process in each channel, and a block of lines defining systematic uncertainties that affect those pdfs.

The pdf can be parametric or template-based, depending on the inputs provided by the user.
In the former case, the parametric pdf for each process has to be provided as a \RooFit object that is derived from the \ttseq{RooAbsPdf} class.
These objects must be contained in a \ttseq{RooWorkspace} object that is identified as an input workspace in the datacard.  
In the latter case, for each channel, histograms must be provided to represent the pdf for each process binned in the observable for that channel.
These must be either \Root \texttt{TH1} or \ttseq{RooDataHist} histogram objects, for analyses in which the data are {\em binned}, or \ttseq{RooDataSet} objects when the data are {\em unbinned}.

As with the counting experiment, the total reference rate of a given process must be identified in the \texttt{rate} line of the datacard.
However, there are special options for shape-based analyses:
\begin{itemize}
  \item A value of \texttt{-1} in the rate line indicates that \Combine should calculate the rate from the input \texttt{TH1} object using the \ttseq{TH1::Integral} method, or the \ttseq{RooDataSet} or \ttseq{RooDataHist} using the \ttseq{RooAbsData::sumEntries} method.
  \item For parametric shapes defined as \ttseq{RooAbsPdf} objects, if a parameter is found in the input workspace with the name \ttseq{pdfname\_norm}, the rate is multiplied by the value of that parameter.
\end{itemize}
For shape analyses, the statistical model constructed by \combine is factorized where possible into normalization and shape terms that provide significant gains in computation time. 
Examples of this factorization can be seen in Eqs.~(\ref{eqn:lambda_binned}) and~(\ref{eqn:parametric-shape-model}).

\subsubsection{Template-based shape analyses}
\label{sec:template-based}
The majority of statistical analyses performed by the CMS Collaboration are template-based analyses. 
This choice of analysis is particularly common when there is no physically motivated parametric function to describe the pdfs for the primary observables.  
Example analyses include the observations of Higgs boson decays to bottom quarks~\cite{CMS:2018nsn}, which uses the output of a deep neural network, and the production of four top quarks in proton-proton collisions~\cite{CMS:2023ftu}, which uses  boosted decision trees to construct the observable.  

A template-based shape analysis is one in which the observable in each channel is partitioned into $N_{B}$ bins. 
The number of events $n_{b}$ in the data that fall within each bin $b$ (with $b$ running from 1 to ${N_{B}}$) is considered as an independent Poisson process. 
The observable $x$ in Eq.~(\ref{eqn:channel-model}) in each channel is replaced by the set of observables $n_{b}$ for the purpose of constructing the statistical model in \combine. 
For each channel, \combine constructs the term $p(x;\vec{\mu},\vtheta)$ as a product of Poisson probabilities, yielding
\begin{linenomath*}
\begin{equation}\label{eqn:pdfbinned}
  p(x;\vec{\mu},\vtheta) =  \prod_{b = 1}^{N_{B}}\mathcal{P}\left({n}_{b}; \lambda_{b}(\vec{\mu}, \vtheta)\right),
\end{equation}
\end{linenomath*}
where the total expectation for a given bin is denoted as $\lambda_{b}$, and the observed event count in data in a given channel and bin is denoted by $n_{b}$.
In the case of $N_{B}=1$, Eq.~(\ref{eqn:pdfbinned}) reduces to the pdf that \combine constructs for a single counting analysis, cf.~Eq.~(\ref{eqn:lh-model2}).
For each channel, histograms must be provided that specify, for each bin, the observed data and the expected yield for each process that contributes.
These can be provided as either \Root \texttt{TH1} objects or \Roofit \ttseq{RooDataHist} objects.
Within each channel, all histograms must use the same partitioning (binning) of the observable.
When using \ttseq{RooDataHist} objects, the  observable name must be the same for all processes within that channel.
An explicit check is made to ensure the normalization of the data histogram corresponds to the number of observed events and that the normalization of the histograms matches the rates provided in the datacard in each channel.

Template-based analysis datacards contain one or more rows in the form:
\begin{lstlisting}[style=mystyle]
shapes <process> <channel> <file> <histogram> [<histogram_systematic_uncertainty_variation>]
\end{lstlisting}

In this datacard line, \ttseq{process} is any of the process names or ``\texttt{*}'' for all processes, or \texttt{data\_obs} for the observed data, and \ttseq{channel} is any of the channel names, or ``\texttt{*}'' for all channels.
The value of \ttseq{file} gives the name of the \Root file. The labels \ttseq{histogram} and 
\texttt{\seqsplit{histogram\_systematic\_uncertainty\_variation}} 
identify either the names of the \texttt{TH1} objects or the \ttseq{RooWorkspace} and \ttseq{RooDataHist} objects within that file.
Several keywords in the datacard line are reserved for automatic substitution when constructing the statistical model:
\begin{itemize}
    \item   \texttt{\$PROCESS} is substituted with each process label or \ttseq{data\_obs} for the observed data.
    \item   \texttt{\$CHANNEL} is substituted with each channel label.
    \item   \texttt{\$SYSTEMATIC} is substituted with the name of each nuisance parameter (with an additional  suffix \texttt{Up} or \ttseq{Down}) that appears in the lines at the end of the datacard.
    \item   \texttt{\$MASS} is replaced with a mass value, which is passed as a command line option. 
\end{itemize}
These lines are interpreted sequentially in the order they appear in the datacard, which allows for multiple instructions with wild card characters and keywords to define the pdfs for each process across the channels.
The \texttt{\$MASS} keyword was originally intended to allow for a single datacard to be used for different mass hypotheses of the Higgs boson $m_{\PH}$. 
In analyses such as the search for $\PH\to\PW\PW^{\ast}\to 2\ell2\PGn$~\cite{CMS:2012zbs}, a different set of histograms for the signal at specific values of $m_{\PH}$ are chosen by specifying this keyword on the command line. In the $\PH\to\PZ\PZ^{\ast}\to 4\ell$ and $\PH\to\PGg\PGg$ analyses, this keyword makes it possible to measure $m_{\PH}$ using the \texttt{MH} parameter, which can be varied continuously~\cite{ATLAS:2015yey}. 
This keyword is useful in searches for new particles whose masses are not known. 
The \combine package supports arbitrary user-defined keywords in the datacard, the values of which can be set at runtime with the command line. 
This is particularly useful for analyses in which the probability distributions depend on more than one  parameter.  

For each channel, the expected value for the total rate in a given bin $b$ is expressed as a sum over processes $p$,
\begin{linenomath*}
  \begin{equation}\label{eqn:lambda_binned}
    \lambda_{b}\left(\vec{\mu},\vtheta\right) =  \sum_{p} M_{p}\left(\vec{\mu},\vtheta\right)\yield_{bp}\left(\vtheta\right) + E_{b}\left(\vec{\mu},\vtheta\right),
  \end{equation}
\end{linenomath*}
where the functions $\yield_{bp}(\vtheta)$ are the number of expected events for a process in a given bin $b$, 
$M_{p}\left(\vec{\mu},\vtheta\right)$ is the overall multiplicative factor representing the effects of the statistical model parameters on the total rate of a given process, and $E_{b}(\vec{\mu},\vtheta)$, discussed at the end of this section, accounts for the statistical uncertainties arising from limited simulated or collision data used to populate the histogram. 
The value of $\lambda_{b}\left(\vec{\mu},\vtheta\right)$ is restricted to be positive for any values of the parameters. 

The functions $\yield_{bp}(\vtheta)$ are derived using the histograms provided by the user that represent the expected distribution for a given process, for the nominal model and for alternates in which a particular nuisance parameter is varied. 
Dropping the process labels henceforth, the nominal bin contents are denoted by $\yield_{b}^{0}$.
For each shape systematic uncertainty $s$, two additional histograms are specified, $\yield_{b}^{s,+}$ and $\yield_{b}^{s,-}$, typically corresponding to the distributions for the values $\anuisance_{s}=+1$ and $\anuisance_{s}=-1$, respectively. 
The datacard must include an additional line in the systematic uncertainties block with the form:
\begin{lstlisting}[style=mystyle]
<name> shape[N] <effect_p_i>
\end{lstlisting}
where \ttseq{effect\_p\_i} is a sequence of values that indicate the effect of the uncertainty in each process $p$ and channel $i$ in the datacard.
Each value in the sequence can be ``\texttt{-}'' or 0 for no effect, or 1 to indicate that the process is affected.
Values different from 1 can be specified to indicate that the histograms provided represent the expected distribution for other values of the associated nuisance parameter $\anuisance_{s}$.
For example, a value of 0.5 would indicate to \combine that the alternative histograms correspond to values of $\anuisance_{s}=\pm 2$ as is illustrated for the systematic uncertainty modelled by the nuisance parameter \ttseq{sigma} in Datacard~\ref{dc:template}.

The alternative histograms can be different from the nominal histogram both in the fractional content of each bin, $f_{b}=\yield_{b}/\sum_{b'}\yield_{b'}$, and in their normalization.
For the latter, the term $M_{p}\left(\vec{\mu},\vec{\nu}\right)$ in Eq.~(\ref{eqn:lambda_binned}) contains an additional asymmetric log-normal multiplicative factor, as described in Table~\ref{tab:countingsys}, where the values of $\kappa^{\text{Up}}$ and $\kappa^{\text{Down}}$ are defined by
\begin{linenomath*}
\begin{equation}
\kappa^{\text{Up/Down}} = \frac{\sum_{b}\yield_{b}^{+/-}}{\sum_{b}\yield_{b}^{0}}.
\end{equation}
\end{linenomath*}
For the former, the user can choose between either parameterizing the variation due to each nuisance parameter directly in terms of the fractional bin contents, $f_b$, or their logarithms, $\ln(f_b)$, by specifying either \ttseq{shape} or \ttseq{shapeN} in the systematic line of the datacard, respectively.
The functions for the fractions are given by
\begin{linenomath*}
\begin{equation}\label{eqn:yield-scale-lin}
  f_{b}(\vtheta) = f^{0}_{b} + \sum_{s} F(\anuisance_{s}, \delta^{s,+}_{b}, \delta^{s,-}_{b}, \epsilon_{s}),
\end{equation}
\end{linenomath*}
if using the \ttseq{shape} algorithm or
\begin{linenomath*}
\begin{equation}\label{eqn:yield-scale-log}
  f_{b}(\vtheta) = \re^{\left(\ln(f^{0}_{b})  + \sum_{s} F(\anuisance_{s}, \Delta^{s,+}_{b}, \Delta^{s,-}_{b}, \epsilon_{s}) \right)}, 
\end{equation}
\end{linenomath*}
if using the \ttseq{shapeN} algorithm, respectively,
where $f^{0}_{b} = \yield^{0}_{b}/\sum_{b'}\yield_{b'}^{0}$, $\delta^{s,\pm}_{b} = f^{s,\pm}_{b} - f^{0}_{b}$, $\Delta^{s,\pm}_{b} = \ln(f^{s,\pm}_{b}) - \ln(f^{0}_{b})$, and the subscript $s$ iterates over each systematic uncertainty.
Similarly to the nominal fractions, the values of $f^{s,\pm}_{b}$ are determined by $f^{s,\pm}_b = \yield^{s,\pm}_{b}/\sum_{b'}\yield^{s,\pm}_{b'}$. 
The \ttseq{shape} algorithm is typically used when the variation due to a systematic uncertainty is relatively small compared to the contents in each bin, while the \ttseq{shapeN} algorithm is used when a systematic uncertainties results in large variations. This is due to the fact that the latter yields a smooth function close to zero when the fractional bin contents are small. 

The function $F$ depends on a set of scaling factors, $\epsilon_{s}$.
These are assumed to be unity by default, but may be set to different values, \eg, if the values of $\yield_{b}^{s,\pm}$ correspond to $\anuisance_{s}= \pm X$, then $\epsilon_{s} = 1/X$.
The function ${F}$ is defined as
\begin{linenomath*}
\ifthenelse{\boolean{cms@external}} {
  \begin{multline}
    F(\anuisance,\delta^{+},\delta^{-},\epsilon)  \\ = 
    \begin{cases}
    \frac{1}{2}\anuisance^{\prime} \bigl( (\delta^{+}-\delta^{-})  \\
    \ \ + \frac{1}{8}(\delta^{+}+\delta^{-})J(\overline{\anuisance}) \bigr), &  -q < \anuisance^{\prime} < q; \\
    \anuisance^{\prime}\delta^{+}, &  \anuisance^{\prime} \ge q;\\
    -\anuisance^{\prime}\delta^{-}, &  \anuisance^{\prime} \le -q;\\
    \end{cases}
  \end{multline}
}{
\begin{equation}
F(\anuisance,\delta^{+},\delta^{-},\epsilon) =
    \begin{cases}
    \frac{1}{2}\anuisance^{\prime} \left( (\delta^{+}-\delta^{-}) + \frac{1}{8}(\delta^{+}+\delta^{-})J(\overline{\anuisance}) \right), &  -q < \anuisance^{\prime} < q; \\
    \anuisance^{\prime}\delta^{+}, &  \anuisance^{\prime} \ge q;\\
    -\anuisance^{\prime}\delta^{-}, &  \anuisance^{\prime} \le -q;\\
    \end{cases}
\end{equation}
}
\end{linenomath*}
where $\anuisance^{\prime} = \anuisance\epsilon$, $\overline{\anuisance} = \anuisance^{\prime} / q$, and $J(\overline{\anuisance})=(3\overline{\anuisance}^5 - 10\overline{\anuisance}^3 + 15\overline{\anuisance})$. 
The minimum value of $\epsilon$ for a given process in a given channel is $q = \min({{\epsilon_{s}}})$. 
These functions ensure that the fractions and their first and second derivatives are continuous for all values of $\anuisance$. A discussion of the different functions that are commonly used for parameterization in template-based analyses can be found in Ref.~\cite{Cranmer:2014lly}.
For each \ttseq{shape[N]} systematic uncertainty line in the datacard corresponding to an uncertainty $s$, an auxiliary observable $\ttheta_{s}$ and its probability distribution $p(\ttheta_{s};\anuisance_{s})=\mathcal{N}(\ttheta_{s};\anuisance_{s},1)$ are included in the statistical model in Eq.~(\ref{eqn:stat-model}).
The default values are $\anuisance_{s}=\ttheta_{s}=0$.

Datacard~\ref{dc:template} is an example of a template-based analysis in which the observable $x$ is the output of a multivariate analysis (MVA) discriminator. 
The histograms defining the distributions of all processes, and their variations due to systematic uncertainties, are contained in a single \Root file named \ttseq{template-analysis-datacard-input.root}. 
Datacard~\ref{dc:template} contains two processes, \ttseq{signal} and \ttseq{background}. 
\begin{figure*}[t]  
\begin{lstlisting}[style=datacard, label=dc:template, caption=Template analysis datacard - \texttt{datacard-2-template-analysis.txt}\vspace{0.25\baselineskip}]
imax 1
jmax 1
kmax 4
# ---------------
shapes * * template-analysis-datacard-input.root $PROCESS $PROCESS_$SYSTEMATIC
# ---------------
bin         ch1
observation 85
# ------------------------------
bin             ch1        ch1
process         signal     background
process         0          1
rate            24         100
# --------------------------------
lumi     lnN    1.1       1.0
bgnorm   lnN    -         1.3
alpha  shape    -          1   # uncertainty in the background template.
sigma  shape    0.5        -   # uncertainty in the signal template.
\end{lstlisting}
\end{figure*}
Line 5 specifies the mapping between the histograms in the \Root file (as shown in Table~\ref{tab:template-contents}) and the processes defined in the datacard.
Lines 17 and 18 provide the definition of systematic uncertainties that affect the probability distribution of the observable for the different processes.
In the example, two such uncertainties are listed, one, \ttseq{alpha}, that affects only the background process, and the other, \ttseq{sigma}, that affects only the signal process.
Any text beyond the number of columns expected in these lines is ignored by \combine and can be used to include descriptions of the systematic uncertainties in the datacard.
Within the \Root file, two histograms for each systematic uncertainty are provided, which represent the probability distributions and rates of the signal and background processes when the associated nuisance parameters are varied.
In the \Root file, these are \texttt{TH1} objects named \ttseq{background\_alphaUp} and \ttseq{background\_alphaDown}, and \ttseq{signal\_sigmaUp} and \ttseq{signal\_sigmaDown}. Table~\ref{tab:template-contents} shows the \texttt{TH1} inputs contained in the \Root file used in Datacard~\ref{dc:template}. 

\begin{table*}[ht]
  \topcaption{\texttt{TH1} objects contained in the \texttt{template-analysis-datacard-input.root} file used in Datacard~\ref{dc:template}. The histograms used to determine the effects of the \texttt{sigma} parameter on the signal distribution correspond to twice the variation resulting from the systematic uncertainty modelled by the parameter.} 
  \centering
  \renewcommand{\arraystretch}{1.6}
  \cmsTable{
    \begin{tabularx}{\textwidth}{>{\hsize=0.35\hsize}X >{\hsize=.65\hsize}X}
      {Object name} & {Description}\\
      \hline
      \texttt{data\_obs}  & Histogram containing the observed number of events in each bin of the analysis.  \\
      \texttt{signal}     & Histogram containing the expected yields $\yield_{b}^{0}$ of the signal process in each bin of the analysis. \\
      \texttt{background} & Histogram containing the expected yields $\yield_{b}^{0}$ of the background process in each bin of the analysis. \\
      \texttt{signal\_sigmaUp},  \texttt{signal\_sigmaDown} & Histograms containing the expected yields of the signal process for each bin in the analysis when the nuisance parameter \texttt{sigma} is set to $-2$ (Down) $\yield_{b}^{-}$, and $+2$ (Up) $\yield_{b}^{+}$, and all other nuisance parameters are set to their default values. \\
      \texttt{background\_alphaUp}, \texttt{background\_alphaDown} & Histograms containing the expected yields of the background process for each bin in the analysis when the nuisance parameter \texttt{alpha} is set to $-1$ (Down) $\yield_{b}^{-}$, and $+1$ (Up) $\yield_{b}^{+}$, and all other nuisance parameters are set to their default values. \\
      \end{tabularx}
  }
  \label{tab:template-contents}
\end{table*}

\begin{figure}[htbp]
  \centering
  \includegraphics[width=\cmsFigWidth]{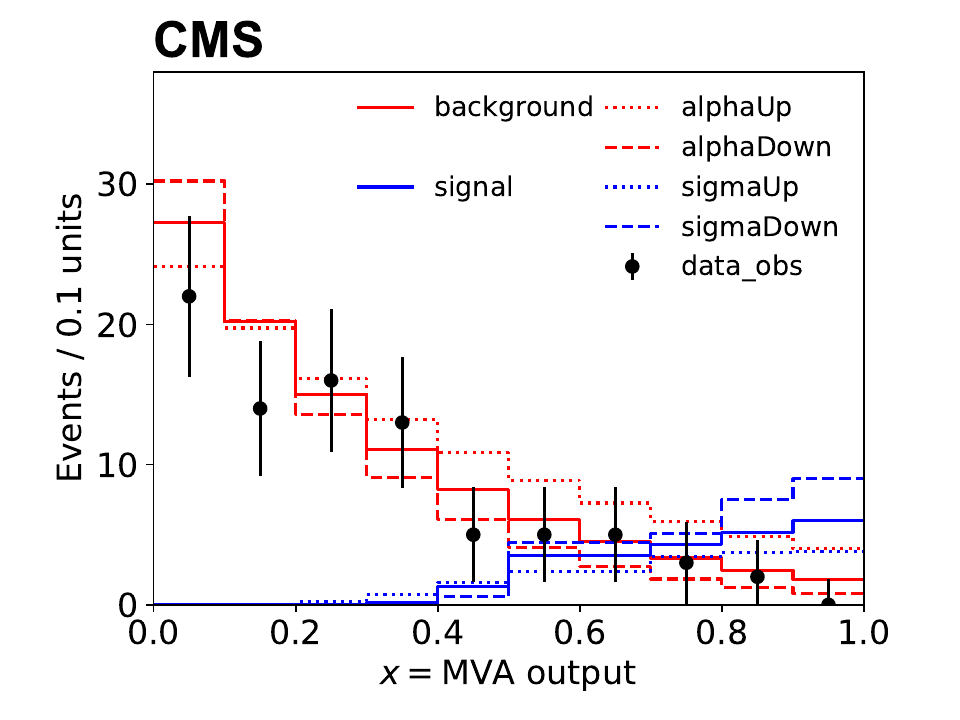}
  \caption{
    Histograms used to define the pdfs for Datacard~\ref{dc:template}. The red and blue histograms show the nominal yields in each bin $\yield_{b}^{0}$ for the background and signal processes, respectively. The dotted and dashed lines show the histograms that provide the values of $\yield_{b}^{+}$ and $\yield_{b}^{-}$, respectively for each of the systematic uncertainties that modify the shape of the signal and background pdfs. The red dashed and dotted lines are associated with the effect of the nuisance parameter \texttt{alpha} on the background process, while the blue dashed and dotted lines 
    are associated with the effect of the nuisance parameter \texttt{sigma} affecting the signal process. The black points show the observed number of events in data in each bin. The error bars indicate  the statistical uncertainty. 
    }
  \label{fig:template-figure}
\end{figure}

Figure~\ref{fig:template-figure} shows the histograms used to determine the expected yield $\yield_{b}^{0}$ and yields expected for the variations of the systematic uncertainties $\yield_{b}^{+}$ and  $\yield_{b}^{-}$ for the signal and background processes.  
The histogram for the observed data is also shown. 

The term $E_{b}\left(\vec{\mu},\vtheta\right)$ in Eq.~(\ref{eqn:lambda_binned}) accounts for the statistical uncertainties in the histograms used to determine $\yield_{bp}$.
When the histograms are provided using \texttt{TH1} objects, these terms can be included using the following line at the end of the datacard:
\begin{lstlisting}[style=mystyle]
<channel> autoMCStats <threshold>
\end{lstlisting}

The first string \ttseq{channel} should give the name of the channels in the datacard for which these statistical uncertainties should be included.
The wildcard ``\texttt{*}'' indicates that this should apply to all channels in the datacard.
The value of \ttseq{threshold} should be set to a value greater than or equal to zero to include the uncertainties.
This value sets the threshold on the effective number of unweighted events above which the uncertainty is modeled with a single nuisance parameter for each bin following the Barlow--Beeston procedure outlined in Ref.~\cite{BARLOW1993219}, using the simplifying approximation introduced in Ref.~\cite{Conway:2011in}. 
This procedure is used to improve the computational performance without introducing significant impact on the accuracy of the results. 
Below the threshold, an individual nuisance parameter for each process is created. 
This threshold improves the computational stability of subsequent statistical routines by reducing the total number of parameters of the statistical model, while maintaining a reasonably accurate description of the statistical uncertainties in the histograms. 
For each nuisance parameter, a corresponding probability distribution for $\ttheta$ is included when building the model in Eq.~(\ref{eqn:stat-model}), the form of which is determined by this effective number.

When \ttseq{threshold} is set to a number of effective unweighted events greater than or equal to zero, denoted $n^\text{threshold}$, the following algorithm is applied to each bin $b$:
\begin{enumerate}
 \item Sum the yields $\yield^{0}_{bp}$ and uncertainties $e_{bp}$ of each background process $p$ in the bin.
$\yield_{\text{tot}} = \sum_{p\,\in\,\text{bkg}}\yield^{0}_{bp}$, and $e_{b,\text{tot}} = \sqrt{\sum_{p\,\in\,\text{bkg}}\,e_{bp}^{2}}$. The values of $e_{ bp}$ are obtained using the 
\texttt{\seqsplit{TH1::GetBinError}} 
method.
 \item If $e_{b,\text{tot}} = 0$, the bin is skipped and no associated nuisance parameters are created.
 \item The effective number of unweighted events is defined as $\yield_{b,\text{tot}}^{\text{eff}} = \yield_{b,\text{tot}}^{2} / e_{b,\text{tot}}^{2}$, rounded to the nearest integer. 
\end{enumerate}
The value of $\yield_{b,\text{tot}}^{\text{eff}}$ determines the functional form of $E_{b}(\vec{\mu},\anuisance)$.
If $\yield_{b,\text{tot}}^{\text{eff}} > n^{\text{threshold}}$, a single auxiliary observable $\ttheta$ is included in the statistical model that is normally  distributed with $p(\ttheta;\anuisance) = \mathcal{N}(\ttheta;\anuisance,1)$. 
The nuisance parameter $\anuisance$ determines the value of $E_{b}$,
\begin{linenomath*}
 \begin{equation}
  E_{b}(\vec{\mu},\vtheta,\anuisance) = \anuisance\left(\sum_{p} e_{bp}^{2}M_{p}^{2}(\vec{\mu},\vtheta)\right)^{\frac{1}{2}},
 \end{equation}
\end{linenomath*}
where $\vtheta$ here represents all of the other nuisance parameters in the statistical model. 
The default values in \combine are $\anuisance=\ttheta=0$. 
If instead $\yield_{b,\text{tot}}^{\text{eff}} \leq n^{\text{threshold}}$, a vector of auxiliary observables $\tvtheta$ with one entry per process is included in the statistical model. 
For processes that have a number of effective events $\yield_{bp}^{2}/e_{pb}^{2}$ less than $n^{\text{threshold}}$, the observable is Poisson distributed with $p(\ttheta_{\alpha};\anuisance_{\alpha}) = \mathcal{P}(\ttheta_{\alpha};\anuisance_{\alpha})$. 
For processes with $w_{bp}^{2}/e_{pb}^{2}\geq n^{\text{threshold}}$, the observable is normally distributed with $p(\ttheta_{\beta};\anuisance_{\beta}) = \mathcal{N}(\ttheta_{\beta};\anuisance_{\beta},1)$. The nuisance parameters determine the value of $E_{b}$, 
\begin{linenomath*}
\ifthenelse{\boolean{cms@external}} {
  \begin{equation}
  \begin{aligned}
    E_{b}(\vec{\mu},\vtheta,\vtheta_{\alpha},\vtheta_{\beta}) & = 
    \sum_{\alpha} \left(\frac{\anuisance_{\alpha}}{\ttheta_{\alpha}} - 1\right)\yield_{ b \alpha}M_{\alpha}(\vec{\mu},\vtheta) \\
    & + \sum_{\beta} \anuisance_{\beta}e_{ b \beta }M_{\beta}(\vec{\mu},\vtheta),
  \end{aligned}
  \end{equation}
}{
\begin{equation}
    E_{b}(\vec{\mu},\vtheta,\vtheta_{\alpha},\vtheta_{\beta}) = \sum_{\alpha} \left(\frac{\anuisance_{\alpha}}{\ttheta_{\alpha}} - 1\right)\yield_{ b \alpha}M_{\alpha}(\vec{\mu},\vtheta)
    + \sum_{\beta} \anuisance_{\beta}e_{ b \beta }M_{\beta}(\vec{\mu},\vtheta),
\end{equation}
}
\end{linenomath*}
where the indices $\alpha$ and $\beta$ run over processes for which the auxiliary observables are Poisson and normally distributed, respectively. 
The default values of $\anuisance_{\beta}=\ttheta_{\beta}$ are zero, while the default values of $\ttheta_{\alpha}$ and $\anuisance_{\alpha}$ are set to the effective number of unweighted events in the \ttseq{TH1} histogram objects used to evaluate $\yield_{bp}^{0}$. 

\subsubsection{Parametric shape analysis}
A parametric shape analysis is one that uses analytic functions rather than histograms to describe the pdfs of continuous primary observables. 
In these cases, the primary observable $x$ in each channel can be univariate or multivariate. 
For example, in the measurements of Higgs boson cross sections in the four-lepton decay mode, the primary observable is bivariate composed of the invariant mass of the four leptons and a kinematic discriminator designed to separate the signal and background processes~\cite{CMS:2021ugl}.
The data in parametric shape analyses can be binned, as in the case of template-based analyses, or unbinned.
Uncertainties affecting the expected distributions of the signal and background processes can be implemented directly as uncertainties in the parameters of those analytical functions.

Datacard~\ref{dc:param} defines a parametric analysis with a single channel and two processes: one signal process and one background process.   
The datacard is similar to what would be used in a search for a narrow resonance over a smooth background, such as in the search for Higgs boson decays to muon pairs~\cite{CMS:2020xwi}.  
The primary observable is the invariant mass $m$ of the decay products, and the signal distribution depends on the hypothesized mass of the resonance, $m_{\PH}$. 
The systematic uncertainties include uncertainties that affect the expected rates of the signal and background processes, and uncertainties in the parameters that describe the signal and background pdfs. 

For datacards describing parametric-based analyses, the term $p(\vec{x};\vec{\mu},\vtheta)$ in Eq.~(\ref{eqn:stat-model}) is constructed in \Combine as,
  \begin{equation}\label{eqn:parametric-shape-model}
    p(\vec{x};\vec{\mu},\vtheta) =  
    \sum_{p}\frac{\lambda_{p}(\vec{\mu},\vtheta) f_{p}(x;\vec{\mu},\vtheta)}{\sum_{p}\lambda_{p}(\vec{\mu},\vtheta)},
\end{equation}
where $f_{p}(x;\vec{\mu},\vtheta)$ are the pdfs for each process $p$ for a given channel with primary observable $x$. 
These pdfs can depend on both the parameters of interest and the nuisance parameters. 
In parametric analyses, for a specific data set with $n$ entries $\left\{x_{d}\right\}$ where $d$ runs from 1 to $n$,  a Poisson probability $\mathcal{P}\left(n;\sum_{p}\lambda_{p}(\vec{\mu},\vtheta)\right)$ is included in the likelihood function in Eq.~(\ref{eqn:lh-model}). 
The Poisson parameters $\lambda_{p}$ depend only on the parameters of interest and any nuisance parameters affecting the rate of a given process.
These parameters are products of multiplicative factors of the form given in Table~\ref{tab:countingsys} and any \ttseq{RooAbsReal} object named \ttseq{pdfname\_norm} found in the input \ttseq{RooWorkspace}, as described in Section~\ref{sec:shapeana}.
When the data are provided as binned data sets with \ttseq{RooDataHist} objects, the continuous observable $x$ is replaced by a sequence of discrete bin centres $x_{b}$, and the pdfs are evaluated at these values. 
If the bins are relatively narrow, this approximation provides a good estimate of the probability density.
When this approximation is not accurate, the \combine package provides a custom class named \ttseq{RooParametricShapeBinPdf} that can wrap any univariate \ttseq{RooAbsPdf} object such that
\begin{linenomath*} 
\begin{equation}
  f_{p}(x_{b};\vec{\mu},\vtheta) = \int_{\text{bin}~b} f_{p}(x;\vec{\mu},\vtheta)\,\rd{x}.
\end{equation}
\end{linenomath*}

The datacard lines for parametric shape analyses need two names to identify the \Roofit object representing the pdf for a given process in each channel, separated by a colon in the following format:
\begin{lstlisting}[style=mystyle]
shapes <process> <channel> <file> <workspace_name>:<pdf_name>
\end{lstlisting}
The label \ttseq{workspace\_name} identifies the input workspace, which is a \ttseq{RooWorkspace} object containing the \Roofit objects, while the second label \ttseq{pdf\_name} identifies the \ttseq{RooAbsPdf} or \ttseq{RooAbsData} contained therein.
The pdfs $f_{p}$ for each process $p$ are defined by the objects identified with \ttseq{pdf\_name}.

\begin{figure*}[t]
\begin{lstlisting}[style=datacard,label=dc:param, caption=Parametric analysis datacard - \texttt{datacard-3-parametric-analysis.txt}\vspace{0.25\baselineskip}]
imax 1
jmax 1
kmax 2
# ---------------
shapes data_obs   bin1 parametric-analysis-datacard-input.root w:data_obs
shapes signal     bin1 parametric-analysis-datacard-input.root w:sig
shapes background bin1 parametric-analysis-datacard-input.root w:bkg
# ---------------
bin                bin1
observation        567
# ---------------
bin                bin1   bin1
process            signal background
process            0      1
rate               10     1
# ---------------
lumi         lnN   1.1    -
sigma        param 1.0    0.1
alpha 	     flatParam
bkg_norm     flatParam
\end{lstlisting}
\end{figure*}
Lines 5--7 indicate the name of the input \Root file 
\texttt{\seqsplit{parametric-analysis-datacard-input.root}} 
that contains an input workspace (\texttt{w}), with \ttseq{RooAbsPdf} objects named \texttt{sig} and \texttt{bkg} defining the pdfs for the signal and background processes, respectively. 
The contents of this workspace are summarized in Table~\ref{tab:param-contents}. 

\begin{table*}[hbt!]
  \topcaption{Contents of the \texttt{RooWorkspace} object contained in the \texttt{parametric-analysis-datacard-input.root} file providing inputs for the parametric analysis datacard.} 
  \centering
  \renewcommand{\arraystretch}{1.6}
  \cmsTable{
    \begin{tabularx}{\textwidth}{llX}
      {Object name} & {Type} & {Description}\\
      \hline
      \texttt{m}        & \texttt{RooRealVar} & The invariant mass observable.  \\

      \texttt{data\_obs}  & \texttt{RooDataSet} & Invariant mass of each event in the observed data.  \\
      \texttt{sig}        & \texttt{RooGaussian} & Normal pdf describing the probability distribution of the invariant mass for the signal process. \\
      \texttt{bkg}        & \texttt{RooExponential} & Exponential pdf describing the probability distribution of the invariant mass for the background process. \\
      \texttt{MH}         & \texttt{RooRealVar}  & Mean of the signal pdf. \\
      \texttt{sigma}      & \texttt{RooRealVar}  & Standard deviation of the signal pdf. \\
      \texttt{alpha}      & \texttt{RooRealVar}  & Slope parameter for the background pdf. \\
      \texttt{bkg\_norm}  & \texttt{RooRealVar}  & Rate multiplier for the total background contribution. \\
      \end{tabularx}
  }
  \label{tab:param-contents}
\end{table*}

There is a single \ttseq{RooRealVar} object named \texttt{m} in the workspace, which represents the primary observable for the analysis. 
The \ttseq{RooDataSet} object named \ttseq{data\_obs} provides the observed data. 
In this datacard the number of events observed in data, as indicated in line 10, is specified as 567, so \combine expects this data set to contain 567 entries, each with its own value of the \ttseq{RooRealVar} object \texttt{m}. 

\begin{figure}[htbp]
  \centering
  \includegraphics[width=\cmsFigWidth]{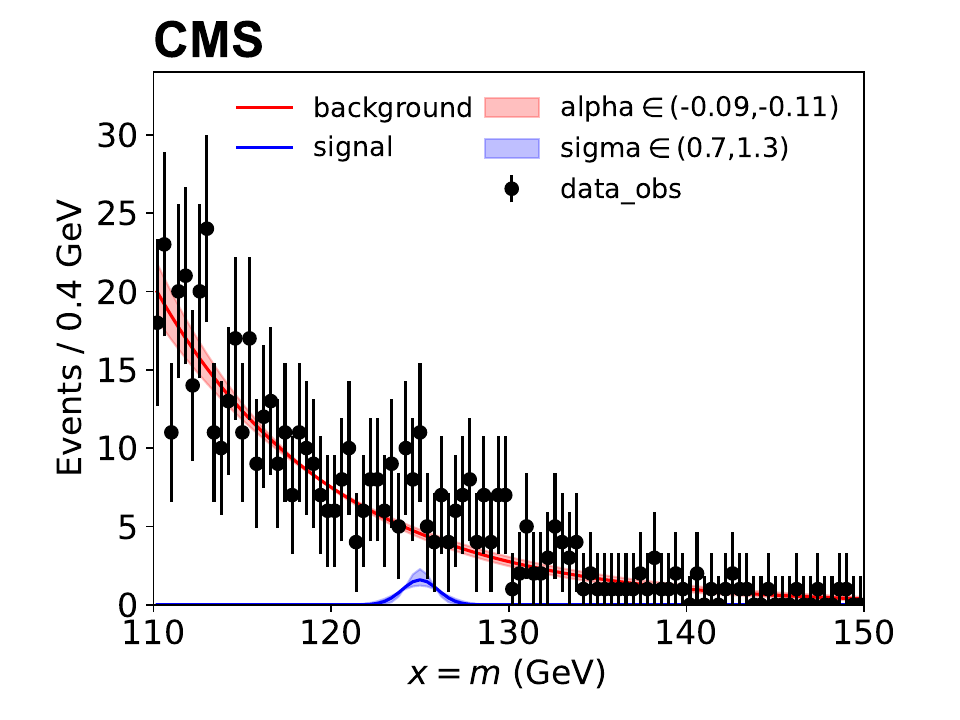}
  \caption{
    Distributions of the invariant mass observable for the signal and background processes defined in Datacard~\ref{dc:param}. The red and blue curves show the parametric functions used to define the probability density for the invariant mass 
    for the background and signal processes, respectively, at the default values of the nuisance parameters, normalized to their expected total yields. The blue shaded band shows the variation of the signal pdf when \texttt{sigma} is varied between 0.7 and 1.3. The red shaded region shows the variation of the 
    background pdf when \texttt{alpha} is varied within 10\% of its default value of $-0.1$. The black points show the distribution of the observed data. The binning and error bars are only for visualization and neither are used by \combine to build the likelihood function. 
    }
  \label{fig:param-figure}
\end{figure}

Figure~\ref{fig:param-figure} shows the pdfs for the signal and background processes, and the distribution of $m$ in observed data. 
The data are unbinned and treated as such in \combine; the binning is performed exclusively for visualization. 
The effects of varying the \ttseq{sigma} and \ttseq{alpha} nuisance parameters on the pdfs for the signal and background processes are  also shown. 

In this datacard, the signal process is parameterized as a normal distribution with a mean given by the hypothesized signal mass value \texttt{MH}.
This variable is used in \combine when interpreting the command line argument value \ttseq{--mass}. 
The value of the hypothesized signal mass will be fixed to the value specified by the \ttseq{--mass} option unless \texttt{MH} is specifically listed as a parameter of interest in the physics model (see Section~\ref{sec:phys_model}). 
The background is an exponential distribution $p(m)\propto \re^{\alpha m}$, with a single nuisance parameter defined in the workspace as a \ttseq{RooRealVar} object named \ttseq{alpha}. 
The workspace also contains a nuisance parameter named \texttt{bkg\_norm} that multiplies the background rate.

Parametric systematic uncertainties are included using datacard lines with the syntax:
\begin{lstlisting}[style=mystyle]
<name> param <V> [<U>, <-UDown/+UUp>]
\end{lstlisting}
These datacard lines directly encode uncertainties in the parameters of the signal and background pdfs.
For each of these lines, an additional term $\mathcal{N}(\ttheta;\anuisance,u)$, is included when constructing the statistical model in Eq.~(\ref{eqn:stat-model}).
The default values for $\ttheta$ and $\anuisance$ are set to the value \texttt{V} and $u$ is the value of \texttt{U} specified in the datacard line.
Line 18 of Datacard~\ref{dc:param} indicates that there is a \texttt{RooRealVar} object describing the parameter \ttseq{sigma}, contained in the workspace, that is associated with a normal distribution with mean specified by \texttt{V}$=$1.0 and standard deviation specified by \texttt{U}$=$0.1.
Asymmetric uncertainties in the parameter can be defined by using the syntax \texttt{-UDown/+UUp}$=-1/{+}1$ standard deviations in the relevant datacard line. 
The corresponding term $p(\ttheta;\anuisance)$ in the statistical model constructed by \combine is a dimidated Gaussian distribution~\cite{Barlow:2003xcj}.
It is possible to include linear correlations between the parameters by first diagonalizing the covariance among them and encoding the resulting linear combinations of parameters as the nuisance parameters that are declared in the datacard. 

To specify that a parameter should be assigned a uniform distribution for $p(\ttheta;\anuisance)$, the datacard line should be:
\begin{lstlisting}[style=mystyle]
<name> flatParam
\end{lstlisting}
The range of allowed parameter values is determined from the \ttseq{RooRealVar} methods \ttseq{getMin} and \ttseq{getMax}, and the default value of the parameter is determined from the \ttseq{getVal} method.
Lines 19 and 20 in Datacard~\ref{dc:param} indicate that there are two such parameters. 
These lines do not count towards the value of \texttt{kmax} since for frequentist calculations they can be dropped from the datacard with no effect on the results. 
The same is not true for Bayesian calculations so it is recommended to include these lines in parametric shape analyses. 

\subsection{Rate parameters}
\label{sec:rate-parameters}
Additional multiplicative scale factors can be introduced in the statistical model that directly modify the rate of a given process, in a given channel, by including additional lines in the datacard for any type of analysis, using the following syntax:
\begin{lstlisting}[style=mystyle]
<name> rateParam <channel> <process> <initial_value> [<min>,<max>]
\end{lstlisting}
A nuisance parameter is included in the statistical model that multiplies the rate of that particular \ttseq{process} in the given \ttseq{channel} by its value.
The default value for this parameter is set to the value indicated by \ttseq{initial\_value}.
The values of \texttt{min} and \texttt{max} can be used to set a range for this parameter.
The same \ttseq{rateParam} nuisance parameter can be attached to multiple channels/processes by using a wild card.
For example, ``\texttt{*}''  matches any process, while ``\ttseq{QCD\_*}'' matches any process whose name begins with ``\texttt{QCD\_}''.
Repeating the same datacard line with different channel/process values is also supported.
A uniform probability distribution within the range given is included in the statistical model if an additional \texttt{flatParam} datacard line is included for that parameter. 

In addition to direct rate modifiers, modifiers that are functions of other parameters can be included using the following syntax:
\begin{lstlisting}[style=mystyle]
<name> rateParam <channel> <process> <formula> <args>
\end{lstlisting}
where \ttseq{formula} is a string with the syntax used by the \Root package's \ttseq{TFormula}, and \texttt{args} is a comma separated list of the arguments for the formula.
Any nuisance parameter can be included in the \ttseq{formula}.

Datacard~\ref{dc:abcd} is an example datacard that uses the \ttseq{rateParam} directive to implement an ABCD background estimation method. 
In the ABCD method, three samples of the data, labeled B, C, and D, that are depleted in signal contributions, are defined using two independent selection variables to estimate one or more background contributions to the signal-enriched data sample labeled A. 
In Ref.~\cite{ABCD}, the ABCD method is used to determine the contribution of the QCD multijet background using the missing transverse momentum and an isolation variable based on the measured energy around the electron in each event. 
The expected contribution from the QCD multijet background in sample A is estimated using the observed yields in samples B, C, and D by assuming the ratio of yields between samples A and B is the same as that between C and D.
An example ABCD analysis can be constructed as a four channel counting analysis in \combine, as described in Datacard~\ref{dc:abcd}: 
\begin{figure*}[t]
\begin{lstlisting}[style=datacard, label=dc:abcd, caption=Example ABCD datacard - \texttt{datacard-4-abcd.txt}\vspace{0.25\baselineskip}]
imax 4  
jmax 1  
kmax *  
# -------
bin         B      C      D      A
observation 50    100     500    16
# -------
bin         B      C      D      A     A
process     bkg    bkg    bkg    bkg   sig
process     1      2      3      4     0
rate        1      1      1      1     3
# -------
lumi   lnN  -      -      -      -     1.02
eff    lnN  -      -      -      -     1.01
alpha rateParam A bkg (@0*@1/@2) beta,gamma,delta
beta  rateParam B bkg 50
gamma rateParam C bkg 100
delta rateParam D bkg 500
\end{lstlisting}
\end{figure*}
The parameters \texttt{beta}, \texttt{gamma}, and \texttt{delta} described by lines 16--18 are simple rate modifiers, $\beta$, $\gamma$, and $\delta$, that directly scale the yields of the \texttt{bkg} process in channels B, C, and D, respectively.
The parameter \texttt{alpha} is determined by the formula $\alpha=\beta \gamma / \delta$, as defined on line 15 of the datacard.  The yield of the \texttt{bkg} process in channel A is scaled accordingly by this formula.

Finally, any pre-existing \ttseq{RooAbsReal} object inside a \Root file containing a \ttseq{RooWorkspace} can be imported into the statistical model using the following syntax:
\begin{lstlisting}[style=mystyle]
<name> rateParam <channel> <process> <rootfile>:<workspacename>
\end{lstlisting}
The value of \texttt{name} should correspond to the name of the \ttseq{RooAbsReal} object inside the 
\texttt{\seqsplit{RooWorkspace}}. 
This allows for arbitrary functions of the statistical model parameters to be used to determine the rate of a particular process in a given channel.

\section{Physics models}
\label{sec:phys_model}
The \combine package supports the construction and association of parameters of interest to the different signal processes declared in the datacard.
This is achieved by defining the parameters of interest and how they affect the signal processes in a Python file: the physics model. 
Different physics models can be used with the same datacard, which facilitates the reinterpretation of a statistical analysis in terms of different parameters of interest~\cite{Cranmer:2021urp}.  
To specify the physics model to use in the statistical model construction, the option \texttt{-P} in \ttseq{text2workspace.py} should indicate the Python file and the model defined therein: 
\begin{lstlisting}[style=commandline]
$ text2workspace.py <datacard.txt> -P HiggsAnalysis.CombinedLimit.<PythonFile>:<modelInstance> [--PO <options>]
\end{lstlisting}
The \ttseq{PythonFile} should be contained in the \ttseq{python/} subdirectory of the \combine package.

The default physics model is one for which the rate of every signal process is multiplied by a common factor $r$, which is the only parameter of interest. 
In this model $\vec{\mu}=r$.
The default model is used if the \texttt{-P} option is not specified.

With the physics model defined, it is now possible to fully determine the statistical model for an input datacard. 
The statistical model created by \combine for Datacard~\ref{dc:counting} when using the default physics model is defined as
\begin{linenomath*}
\ifthenelse{\boolean{cms@external}} {
  \begin{equation}
  \begin{aligned}
    p(n, \tvtheta; r,\vtheta)  & = 
    \dfrac{\lambda(r,\vtheta)^{n}}{n!} \re^{-\lambda(r,\vtheta)} \\
    & \times \frac{1}{2\pi}\re^{-\left(\anuisance_{\texttt{lumi}}-\ttheta_{\texttt{lumi}}\right)^2} \re^{-\left(\anuisance_{\texttt{xs}}-\ttheta_{\texttt{xs}}\right)^2} \\
    & \times
     \frac{\left(\anuisance_{\texttt{nWW}}\right)^{\ttheta_{\texttt{nWW}}}}{\ttheta_{\texttt{nWW}}!}\re^{-\anuisance_{\texttt{nWW}}},
  \end{aligned}
  \end{equation}
}{
  \begin{equation}
    p(n, \tvtheta; r,\vtheta) =
    \dfrac{\lambda(r,\vtheta)^{n}}{n!} \re^{-\lambda(r,\vtheta)} 
    \frac{1}{2\pi}\re^{-\left(\anuisance_{\texttt{lumi}}-\ttheta_{\texttt{lumi}}\right)^2} \re^{-\left(\anuisance_{\texttt{xs}}-\ttheta_{\texttt{xs}}\right)^2}
     \frac{\left(\anuisance_{\texttt{nWW}}\right)^{\ttheta_{\texttt{nWW}}}}{\ttheta_{\texttt{nWW}}!}\re^{-\anuisance_{\texttt{nWW}}},
  \end{equation}
}
\end{linenomath*}
where the function $\lambda(r,\vtheta)$ is given by
\begin{linenomath*}
\ifthenelse{\boolean{cms@external}} {
  \begin{equation}
    \begin{aligned}
    \lambda(r,\vtheta) & =  r \; 1.47 \; (1.11)^{\anuisance_{\texttt{lumi}}} (1.2)^{\anuisance_{\texttt{xs}}}  \\
    & + 0.22  \;(1.11)^{\anuisance_{\texttt{lumi}}}  \\
    & + 0.64 \;(1.11)^{\anuisance_{\texttt{lumi}}}\frac{\anuisance_{\texttt{nWW}}}{4}.
    \end{aligned}
  \end{equation}
}{
  \begin{equation}
    \lambda(r,\vtheta) = r \; 1.47\;(1.11)^{\anuisance_{\texttt{lumi}}} (1.2)^{\anuisance_{\texttt{xs}}} 
    + 0.22  \;(1.11)^{\anuisance_{\texttt{lumi}}} 
    + 0.64 \;(1.11)^{\anuisance_{\texttt{lumi}}}\frac{\anuisance_{\texttt{nWW}}}{4}.
  \end{equation}
}
\end{linenomath*}
The observable values for the data are set to $n=0$, $\ttheta_{\texttt{lumi}}=\ttheta_{\texttt{xs}}=0$, and $\ttheta_{\texttt{nWW}}=4$. 

Generic physics models can be implemented by writing a Python class that defines the parameters of interest and defines how the signal (and background) yields depend on these parameters. 
There are numerous example physics models provided in the \combine package in  \ttseq{python/PhysicsModel.py} and other Python files within the same directory.
In the \ttseq{PhysicsModel:floatingXSHiggs} physics model, the signal processes expected in the datacard correspond to the four dominant Higgs boson  production modes at the LHC: gluon fusion, vector boson fusion, and Higgs boson production associated with a vector boson or a pair of top quarks.
Their rates are modified by separate scaling parameters; \texttt{r\_ggH}, \texttt{r\_qqH}, \texttt{r\_VH} (or \texttt{r\_WH} and \texttt{r\_ZH}), and \texttt{r\_ttH} as defined in the following block of code:
\begin{lstlisting}[style=mystyle]
def doParametersOfInterest(self):
  """Create parameters of interest (POIs) and other parameters, and define the POI set."""
  # --- Signal Strength as only POI ---
  if "ggH" in self.modes: self.modelBuilder.doVar("r_ggH[1,%s,%s]" % (self.ggHRange[0], self.ggHRange[1]))
  if "qqH" in self.modes: self.modelBuilder.doVar("r_qqH[1,%s,%s]" % (self.qqHRange[0], self.qqHRange[1]))
  if "VH"  in self.modes: self.modelBuilder.doVar("r_VH[1,%s,%s]"  % (self.VHRange [0], self.VHRange [1]))
  if "WH"  in self.modes: self.modelBuilder.doVar("r_WH[1,%s,%s]"  % (self.WHRange [0], self.WHRange [1]))
  if "ZH"  in self.modes: self.modelBuilder.doVar("r_ZH[1,%s,%s]"  % (self.ZHRange [0], self.ZHRange [1]))
  if "ttH" in self.modes: self.modelBuilder.doVar("r_ttH[1,%s,%s]" % (self.ttHRange[0], self.ttHRange[1]))
  poi = ",".join(["r_"+m for m in self.modes])
  if self.pois: poi = self.pois
  ...
\end{lstlisting}
Each of these is a parameter of interest in the statistical model that \combine constructs. 
The arrays \ttseq{ggHRange}, \ttseq{qqHRange}, \ttseq{VHRange}, \ttseq{WHRange}, \ttseq{ZHRange} and \ttseq{ttHRange} specify the range of each parameter of interest and are defined in the same Python class. 
The association of each parameter of interest with each production process is defined in the following function:
\begin{lstlisting}[style=mystyle]
def getHiggsSignalYieldScale(self,production,decay, energy):
  if production == "ggH": return ("r_ggH" if "ggH" in self.modes else 1)
  if production == "qqH": return ("r_qqH" if "qqH" in self.modes else 1)
  if production in [ "WH", "ZH", "VH" ]: return ("r_VH" if "VH" in self.modes else 1)
  if production == "ttH": return ("r_ttH" if "ttH" in self.modes else ("r_ggH" if self.ttHasggH else 1))
  raise RuntimeError, "Unknown production mode '%s'" % production
\end{lstlisting}
An example datacard with two signal processes and two channels for use with 
\texttt{\seqsplit{PhysicsModel:floatingXSHiggs}} 
is shown in Datacard~\ref{dc:multisig}. 
In this datacard, there are two signal processes, \texttt{ggH\_hgg} and \texttt{qqH\_hgg}, that correspond to Higgs boson production in the gluon fusion and vector boson fusion modes, respectively, decaying to two photons. 
The background process is estimated by fitting the data outside the signal peak. 
The channels correspond to events with an additional pair of jets reconstructed (\texttt{dijet}) or otherwise, leading to a more inclusive channel (\texttt{incl}).
\begin{figure*}[t]
\begin{lstlisting}[style=datacard,label=dc:multisig,caption=Multi signal datacard - \texttt{datacard-5-multi-signal.txt}\vspace{0.25\baselineskip}]
imax 2
jmax 2
kmax *
------------
shapes *  dijet  FAKE
shapes *  incl   FAKE
------------
bin           incl dijet
observation   166  8
------------
bin               incl    incl    incl    dijet   dijet   dijet
process           ggH_hgg qqH_hgg bkg     ggH_hgg qqH_hgg bkg
process            -1     0       1       -1      0        1
rate              21      1.6     140     0.4     0.95     3.2
------------
QCDscale_ggH lnN  1.12    -       -       1.12    -        -
pdf_gg       lnN  1.08    -       -       1.08    -        -
pdf_qqbar    lnN   -      1.025   -       -       1.025    -
bg_incl      lnN   -      -       1.05    -       -        -
\end{lstlisting}
\end{figure*} 
The \ttseq{FAKE} directive in lines 5 and 6 are used to indicate that each channel of the counting analysis datacard represents a single bin in a histogram. 
This is required in counting analysis datacards to run some of the diagnostic methods described in Section~\ref{goodness-of-fit-tests}, and does not change the statistical model constructed by \combine. 

It is possible to include generic constraints on the parameters of the physics model.
These can be included in the datacard with lines having the following syntax:
\begin{lstlisting}[style=mystyle]
<name> constr <formula> <args> <delta>
\end{lstlisting}
where \ttseq{name} should be a unique identifier for the constraint and \ttseq{formula} and \ttseq{args} follow the \Root \ttseq{TFormula} syntax.
The result is to multiply the probability term $p(\vec{x};\vec{\mu},\vtheta)$ in Eq.~(\ref{eqn:stat-model}) by the product of constraint terms,
\begin{linenomath*}
\begin{equation}
  p(\vec{x};\vec{\mu},\vtheta) \to p({\vec{x};\vec{\mu},\vtheta}) \prod_{l} \frac{1}{\delta_{l}\sqrt{2\pi}}\re^{-\frac{1}{2}\left(\dfrac{g_{l}(\vec{\mu})}{\delta_{l}}\right)^{2}},
\end{equation}
\end{linenomath*}
where $l$ runs over the \ttseq{constr} lines in the datacard. 
This feature can be used to include additional restrictions on the parameters of the model imposed by external theoretical or experimental constraints. 
This feature has been used to perform regularization in measurements of unfolded differential Higgs boson cross sections in the  $\PH\to\PGt\PGt$ decay mode~\cite{CMS:2021gxc}. 
As an example, the following datacard line produces a single constraint term in the statistical model with $g(\vec{\mu}) = \rggH-2r_{\PV\PH}+r_{\PGt\PGt\PH}$ and $\delta=0.03$, when used with the \ttseq{PhysicsModel:floatingXSHiggs} physics model: 
\begin{lstlisting}[style=mystyle]
constraint_higgs constr @0-2*@2+@1 r_ggH,r_VH,r_ttH 0.03
\end{lstlisting}
The order of the list of parameters in the fourth column of the datacard line defines which of the parameters is assigned to each term in $g(\vec{\mu})$.  

Throughout this paper, $\mu$ generically denotes the first parameter of interest defined in the physics model, while $r$ specifically refers to the single parameter of interest in the default physics model.
For any physics model, it is possible to redefine the list of parameters of interest, or their order within the list, using the \combine command line option \ttseq{--redefineSignalPOIs}.
Parameters of interest not included in this list are demoted to nuisance parameters.
This command may include nuisance parameters, which results in the removal of the probability density in the statistical model for the associated auxiliary observable. 
This can be used to test how well any parameter of the model can be measured using only the primary observables, and any remaining auxiliary observables of a given data set.   

\section{How to run \altcombine}
\label{sec:runningthetool}
This section gives an overview of the command line executable \texttt{combine} provided by the package, which is used to perform a number of different statistical routines using the statistical model constructed by \combine.
The executable runs using the command:
\begin{lstlisting}[style=commandline]
$ combine <datacard.[txt|root]> -M <Method>
\end{lstlisting}
Where the \texttt{Method} specifies the statistical calculation to be performed.
A list of available options for the executable is displayed by adding the command line option \ttseq{-{}-help}.

\subsection{Generic minimizer options}\label{sec:generic-minimizer-options}
A number of methods available in \combine make use of numerical optimization of the likelihood function given in Eq.~(\ref{eqn:lh-model}). 
Typically, these  methods make use of the profile negative-log-likelihood function,  $-\ln{\Likelihood(\vec{\mu},\hat{\hat{\vtheta}}(\vec{\mu}))}$, in which the nuisance parameters are profiled; $\hat{\hat{\vtheta}}(\vec{\mu})$ are the values of the nuisance parameters $\nu$ that maximize the likelihood function at a fixed set of values of the parameters of interest $\vec{\mu}$. 
The class \texttt{CascadeMinimizer} is used to steer these routines and allows for a sequential minimization of $-\ln{\Likelihood(\vec{\Phi})}$ using different algorithms.
The class also allows for \combine to perform minimization over any discrete nuisance parameters like those needed for the implementation of the discrete profiling method described in Ref.~\cite{Dauncey:2014xga}.
The combinations of minimizers and algorithms supported in \combine are given in Table~\ref{tab:minimizertypes}.
Details of these algorithms can be found in Refs.~\cite{galassi2018} and~\cite{James:1975dr}.

\begin{table*}[ht!]
\topcaption{Available combinations of minimizer and algorithms in \combine.}
\centering
\cmsTable{
\begin{tabularx}{\textwidth}{llX}
Minimizer & Reference & Available algorithms
\\
\hline
\texttt{Minuit} & ~\cite{James:1975dr} & \texttt{Migrad}, \texttt{Simplex}, \texttt{Combined},
\texttt{Scan}.
\\
\texttt{Minuit2} & ~\cite{James:1975dr} & \texttt{Migrad}, \texttt{Simplex}, \texttt{Combined},
\texttt{Scan}.
\\
\texttt{GSLMultiMin} & ~\cite{galassi2018} &  \texttt{ConjugateFR}, \texttt{ConjugatePR},
\texttt{BFGS}, \texttt{BFGS2}, \texttt{SteepestDescent}.
\\
\end{tabularx}
}
\label{tab:minimizertypes}
\end{table*}

\subsection{Output from \combine}
\label{output-from-combine}

Most of the methods available in \combine output the results of the computation to the terminal.
In addition these results are also saved in a \Root file containing a \ttree called \ttseq{limit}.
The name of this file has the following format:

\begin{lstlisting}[style=mystyle]
higgsCombine$NAME.MethodName.mH$MASS.[$WORD$VALUE].root
\end{lstlisting}
where \texttt{NAME} is set to the value passed to the option \texttt{-n}, which defaults to \texttt{Test}, and \ttseq{\$WORD\$VALUE} is any user-defined keyword \texttt{WORD} in the datacard that has been set to a particular value \texttt{VALUE} using the command line option \ttseq{--keyword-value} \ttseq{WORD=VALUE}. 
The option can be repeated multiple times for multiple keywords. 
The keyword-value pairs are also stored in the output \Root file.
The option \texttt{-m} sets the value of \texttt{\$MASS} and the parameter \texttt{MH} if it is included in the statistical model.
Its value is written to the branch \texttt{mh} in the output \ttree object.

The structure of the \ttree contained in the output \Root file is given in Table~\ref{tab:outputbranches}. 
The branch names \ttseq{limit} and \ttseq{limitErr} always refer to the main result of any statistical routine and an estimate of its uncertainty, whether or not that routine calculates a frequentist limit.

\begin{table*}[ht!]
\topcaption{\texttt{TTree} branches contained in the output \Root file from \combine.}
\centering
\cmsTable{
\begin{tabularx}{\textwidth}{llX}
{Branch name} & {Type} & {Description}
\\
\hline
{\texttt{limit}} & \texttt{Double\_t} & Main result of the statistical routine being performed.
\\
{\texttt{limitErr}} & \texttt{Double\_t} & Estimated uncertainty
in the result.
\\
{\texttt{mh}} & \texttt{Double\_t} & Value 
specified with \ttseq{--mass} command line option. The default value is 120. 
\\
{\texttt{iToy}} & \texttt{Int\_t} & Pseudo-data set identifier if
running with \ttseq{--toys}.
\\
{\texttt{iSeed}} & \texttt{Int\_t} & Random seed specified with
\texttt{-s}.
\\
{\texttt{t\_cpu}} & \texttt{Float\_t} & Estimated processing time.
\\
{\texttt{t\_real}} & \texttt{Float\_t} & Elapsed wall-clock time for routine.
\\
{\texttt{quantileExpected}} & \texttt{Float\_t} & Quantile identifier for methods that calculate expected and observed results. 
The meaning is method-dependent.
Negative values are reserved for entries that are not related to quantiles of a calculation. 
The default is set to $-1$ and specifies that the entry corresponds to the result obtained from the observed data.
\\
\end{tabularx}
}
\label{tab:outputbranches}
\end{table*}

\subsection{Pseudo-data generation}\label{toy-data-generation}
By default, \combine performs the calculation using the observed data. For example, in frequentist methods the observed data are automatically used to construct the likelihood function in Eq.~(\ref{eqn:lh-model}). 
It is possible to run these routines instead using pseudo-data sets to determine the distributions of various statistical quantities such as maximum likelihood estimates, or perform optimization studies that are blind to the observed data.

The option \ttseq{--toys} is used to instruct \combine to first generate one or more pseudo-data sets, which are  used in place of the observed data.
There are two variants of this procedure available in \combine. 
In the first, specified by \texttt{--toys <N>} with $N>0$, \combine generates $N$ pseudo-data sets from the statistical model and runs the specified statistical routine once per data set. 
The pseudo-data set is constructed by generating random values of the observables $\vec{x}$. The random number seed for the generation can be modified with the option \texttt{--seed <value>}, which allows the user to ensure each run of the \combine command produces different pseudo-data sets, or identical pseudo-data sets. 
This allows certain calculations to be split into parallel tasks in the former case and for performing diagnostic studies in particular pseudo-data sets in the latter.
The output \ttree contains one entry for each of these data sets when generated with $N>0$.

In the second variant, specifying \texttt{--toys -1} produces an Asimov data set~\cite{Cowan:2010js}. 
An Asimov data set is defined as that in which the maximum likelihood estimates for all of the model parameters are equal to the values used to generate the data set. 
Asimov data sets are used for deriving the expected outcome of frequentist calculations such as in the determination of upper limits and confidence intervals. 
Where valid, their use makes these calculations much more computationally efficient than using the first variant of pseudo-data generation. 

In \combine,  Asimov data sets are constructed using the expectation value for the probability $p(\vec{x};\vec{\mu},\vtheta)$ in counting analyses and shape analyses for which the data are binned, or by using a large sample of weighted events, which are generated according to $p(\vec{x};\vec{\mu},\vtheta)$. 
In the former case, the values of $n_{b}$ for the Poisson probabilities in Eq.~(\ref{eqn:pdfbinned}) are allowed to be non-integer, since the terms $n_{b}!$ are ignored when constructing a likelihood for the statistical routines available in \combine. 
The event weights in the latter case are identical for every event in the same channel, and are accounted for when estimating parameter uncertainties from Asimov data sets~\cite{Langenbruch:2019nwe}.
The default values of the parameters of interest are used when generating pseudo-data sets in both variants.
The command line option \ttseq{-{}-setParameters <x>=<value\_x>,<y>=<value\_y>,...} can be used to specify other values of the parameters to be used for the generation of the pseudo-data sets.

By default, pseudo-data sets in \combine make use of marginalization where the value of each nuisance parameter $\aknuisance$ is randomly sampled from its probability distribution $p(\aknuisance|\tthetak)$ in Eq.~(\ref{eq:priors}) before generating values for the primary observables in each pseudo-data set. 
The auxiliary observables $\tvtheta$ are set to their default values. 
This can be modified by specifying the option \texttt{-{}-toysFrequentist}. 
With this option and $N>0$, a parametric bootstrap~\cite{Efron1979,Lee2005} is instead performed: each $\ttheta$ is generated according to its probability distribution $p(\ttheta;\anuisance)$, where the value of the nuisance parameter $\anuisance$ is set to the maximum likelihood estimate obtained using the observed data and fixing the parameters of interest to the values specified in the \ttseq{-{}-setParameters} option. 
If instead $N=-1$, the value of $\ttheta$ is this maximum likelihood estimate for the corresponding nuisance parameter $\anuisance$. 
When specifying the option \ttseq{--bypassFrequentistFit}, the default values of the nuisance parameters instead of the maximum likelihood estimates are used.  

It is possible to separate the tasks of generating the pseudo-data sets from running statistical routines, by first saving the pseudo-data sets to a \Root file on disk, and then passing them to any \combine method later.
The following sections describe the most commonly used statistical methods available in \combine.
These represent only part of the functionality of the tool and users are recommended to consult the online documentation for a full description of its capabilities.

\subsection{Frequentist limits and confidence intervals}\label{sec:intervals}
The \texttt{HybridNew} method can be used for calculating upper limits with statistical models created with the default physics model, and for calculating confidence intervals for models with one or more parameters of interest using the Feldman--Cousins procedure~\cite{Feldman:1997qc}. 
Each of these methods utilizes a test statistic based on the likelihood function given in Eq.~(\ref{eqn:lh-model}).

In the case of upper limits, the single parameter of interest $\mu$ corresponds to the parameter $r$ in the default physics model.
A number of prescriptions using different test statistics are supported in \combine as follows:
\begin{itemize}
\itemsep1pt\parskip0pt\parsep0pt
\item
  {LEP-style}:
  \texttt{-{}-testStat LEP} \ttseq{--generateNuisances=1} \ttseq{--fitNuisances=0}.
    The test statistic is defined using the ratio of likelihoods,
    \begin{linenomath*}
    \begin{equation}
    q_{\mathrm{LEP}}(\mu)=-2\ln\left(\frac{\Likelihood(\mu=0,\vtheta_{0})}{\Likelihood(\mu,\vtheta_{0})}\right),
    \end{equation}
  \end{linenomath*}
    where $\vtheta_{0}$ are the default values of the nuisance parameters. This test statistic was used in the searches for the Higgs boson at the LEP~\cite{LEPWorkingGroupforHiggsbosonsearches:2003ing}. 
\item
  {TEV-style}:
  \texttt{--testStat TEV} \ttseq{--generateNuisances=0} 
  \texttt{\seqsplit{--generateExternalMeasurements=1}} 
  \ttseq{--fitNuisances=1}.
    The test statistic is defined using a ratio of profile likelihoods,
    \begin{linenomath*}
    \begin{equation}
    q_{\mathrm{TEV}}(\mu)=-2\ln\left(\frac{\Likelihood(\mu=0,\hat{\hat{\vtheta}}(0))}{\Likelihood(\mu,\hat{\hat{\vtheta}}({\mu}))}\right),
    \end{equation}
  \end{linenomath*}
    where  $\hat{\hat{\vtheta}}({0})$, and $\hat{\hat{\vtheta}}({\mu})$  are the values of the nuisance parameters that maximize the likelihood function at $\mu=0$ and $\mu$ respectively.
    For the purposes of pseudo-data generation, the nuisance parameters are set to the values of $\hat{\hat{\vtheta}}({\mu})$ obtained using the observed data, while the values of $\tvtheta$ are randomly sampled according to their probability densities. This test statistic was used in the context of Higgs boson searches at the Tevatron~\cite{CDF:2013kiv}. 
\item
  {LHC-style}: \texttt{-{}-LHCmode LHC-limits}. 
  The test statistic is defined using a ratio of profile likelihoods,
  \begin{linenomath*}
  \begin{equation}\label{eqn:lhcteststat}
  \widetilde{q}_{\mathrm{LHC}}(\mu) = \begin{cases}
      -2\ln\left(\dfrac{ \Likelihood(\mu,\hat{\hat{\vtheta}}({\mu}))}{ \Likelihood(\hat{\mu},\hat{\vtheta})}\right) & \text{if } 0\leq\hat{\mu} \leq{\mu}, \\
      -2\ln\left(\dfrac{ \Likelihood(\mu,\hat{\hat{\vtheta}}({\mu}))}{ \Likelihood(0,\hat{\hat{\vtheta}}(0))}\right) & \text{if } \hat{\mu} < 0, \\
      0 & \text{if } \hat{\mu} > {\mu},
    \end{cases}
  \end{equation}
\end{linenomath*}
  where $\hat{\mu}$ is the maximum likelihood estimator for $\mu$. The same result is obtained with the option \ifthenelse{\boolean{cms@external}}{\\}{}\texttt{-{}-testStat LHC} \ttseq{-{}-generateNuisances=0} 
  \texttt{\seqsplit{-{}-generateExternalMeasurements=1}} 
  \ttseq{-{}-fitNuisances=1}.
  The values of the nuisance parameters $\vtheta$ that maximize the likelihood $\Likelihood$ assuming a specific value of $\mu$ and for $\mu=\hat{\mu}$ are denoted $\hat{\hat{\vtheta}}({\mu})$ and $\hat{\vtheta}$, respectively. 
\end{itemize}
The test statistic in Eq.~(\ref{eqn:lhcteststat}) is the most widely used for setting upper limits in searches for new physics at the LHC~\cite{LHC-HCG}. 
In \combine, these test statistics can be modified to perform two point hypothesis tests such as those performed to test different hypotheses for the spin and parity of the Higgs boson~\cite{CMS:2014nkk}.
In the LEP-style prescription, the nuisance parameter values for each pseudo-data set are randomly sampled from their distributions $p(\aknuisance|\tthetak)$ in Eq.~(\ref{eq:priors}). 
In the TEV- and LHC-style prescriptions, a parametric bootstrap is used where the nuisance parameters are set to the values of $\hat{\hat{\vtheta}}(\mu)$ obtained using the observed data before generating the primary and auxiliary observables from their probability distributions. 
It is also possible to integrate out (marginalize) the nuisance parameters in a Bayesian-inspired procedure~\cite{intL} and advocated for upper limits in high-energy physics analyses~\cite{COUSINS1992331}. 
This amounts to calculating
\begin{linenomath*}
\begin{equation}
  \mathcal{L}_{\text{int}}(\vec{\mu}) = \int \mathcal{L}(\vec{\Phi})\prod_{k}\pi_{k}(\anuisance_{k})d\vtheta,
\end{equation}
\end{linenomath*}
where $\mathcal{L}_{\text{int}}(\vec{\mu})$ is the integrated likelihood~\cite{intL}. In \combine, this is achieved by modifying the options to \ttseq{-{}-generateNuisances=1} and \ttseq{-{}-generateExternalMeasurements=0}. 
This is required to calculate upper limits in cases where there are few or no background events in channels dominated by signal processes such as in the search for lepton flavor violating tau lepton decays performed by the CMS Collaboration~\cite{CMS-PAS-BPH-21-005}. 

For a specific value of $\mu$, the value of the test statistic using the observed data $q^{\text{obs}}(\mu)$ is calculated, along with the two $p$-values $p_{\mu}$ and $p_{b}$ defined as
\begin{linenomath*}
  \begin{equation}
    p_{\mu} = \begin{cases}
      \int_{\qxobsm}^{\infty} f(q_{\text{x}}(\mu)|\mu)\,\rd{q_{\text{x}}} & \text{if x=LHC}, \\
      \int_{-\infty}^{\qxobsm} f(q_{\text{x}}(\mu)|\mu)\,\rd{q_{\text{x}}} & \text{if x=TEV or LEP},
    \end{cases}
  \end{equation}
\end{linenomath*}
and
\begin{linenomath*}
  \begin{equation}
    p_{b} = \begin{cases}
      \int_{0}^{\qxobsm} f(q_{\text{x}}(\mu)|0)\,\rd{q_{\text{x}}} & \text{if x=LHC}, \\
      \int_{\qxobsm}^{\infty} f(q_{\text{x}}(\mu)|0)\,\rd{q_{\text{x}}} & \text{if x=TEV or LEP},
    \end{cases}
  \end{equation}
\end{linenomath*}
where the distributions of the test statistics $f(q_{\text{x}}(\mu)|\mu)$ and $f(q_{\text{x}}(\mu)|0)$ are determined using pseudo-data sets, assuming the value of $\mu$ indicated and the values of $\vtheta$ and $\tvtheta$ depending on the options used, as described above.
From these $p$-values, the tool calculates the \CLs criterion~\cite{Junk,Read,Workman:2022ynf} defined by $\CLs=p_{\mu}/(1-p_{b})$.
The tool then uses a bisection algorithm to find the value of $\mu$ for which $\CLs=\alpha$ corresponding to the upper limit on $\mu$ at the $100 (1-\alpha)\%$ confidence level (\CL).
The tool can also calculate the median, and 2.5, 16, 85, or 97.5\% quantiles of the expected distribution of the upper limit assuming $\mu=0$, by including the option \ttseq{-{}-expectedFromGrid=<X>}, where \texttt{X} is either 0.5, 0.025, 0.16, 0.84, or 0.975, respectively.

\begin{figure}[htbp]
  \centering
  \includegraphics[width=\cmsFigWidth]{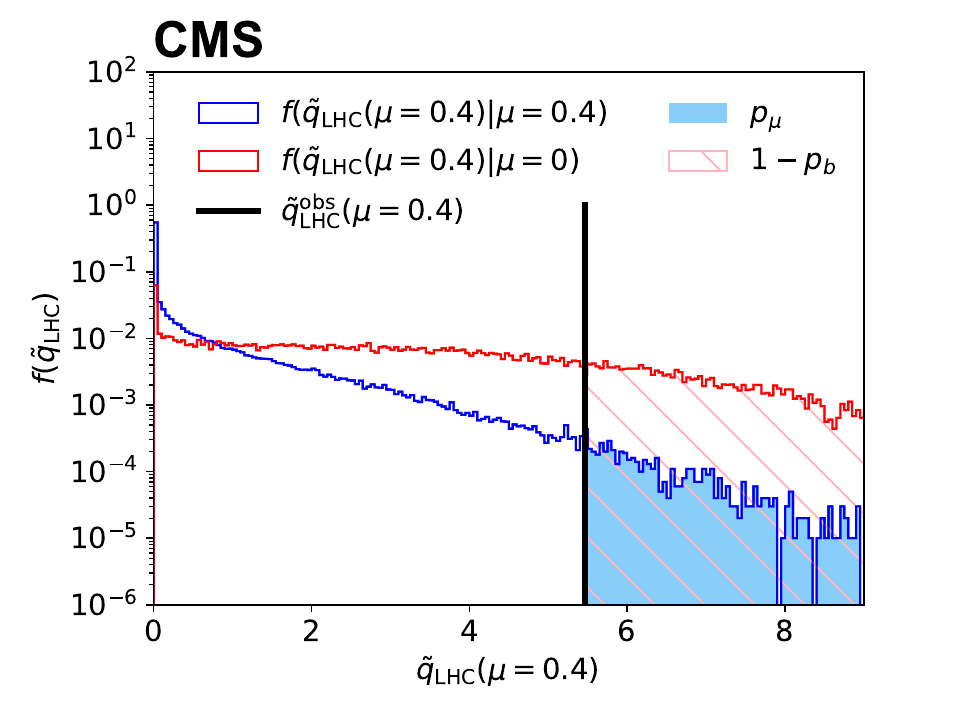}
  \caption{Distributions of $\widetilde{q}_{\mathrm{LHC}}(\mu=0.4)$ from 100,000 pseudo-data sets for $\mu=0$ (red histogram) and $\mu=0.4$ (blue histogram) using the analysis described in Datacard~\ref{dc:template}.
  The observed value of the test statistic is indicated by the black vertical line and the regions used to determine $1-p_{b}$ and $p_{\mu}$ are indicated by the pink hatched and light blue shaded regions, respectively.}
  \label{fig:exampleLHCteststat}
\end{figure}

The 95\% \CL upper limit on $\mu$ in the example template analysis can be calculated using Datacard~\ref{dc:template} with the command: 
\begin{lstlisting}[style=commandline]
$ combine datacard-2-template-analysis.txt -M HybridNew --LHCmode LHC-limits --rMax 2.0 --clsAcc 0.01
\end{lstlisting}
The option \ttseq{--rMax} specifies the maximum value of $\mu$ in the algorithm that performs the search for the upper limit. 
The results of the calculation are output to the terminal as:
\begin{lstlisting}[style=output]
>  -- Hybrid New --
> Limit: r < 0.346362 +/- 0.0134581 @ 95% CL
> Done in 0.31 min (cpu), 0.32 min (real)
\end{lstlisting}
The bisection algorithm for calculating the upper limit terminates when the estimate of the precision on the upper limit value is below a specified threshold, or when the precision cannot be improved further.
The user can specify the following options to control this behavior:
\begin{itemize}
\itemsep1pt\parskip0pt\parsep0pt
\item
  \texttt{--rAbsAcc} and \texttt{--rRelAcc}: Define the accuracy on the upper limit, $\mu_\text{up}$ at which the algorithm terminates. The default values are 0.1 and 0.05
  respectively, meaning that the search terminates when the absolute accuracy $\Delta \mu_\text{up} < 0.1$ or the relative accuracy $\Delta \mu_\text{up}/\mu_\text{up} < 0.05$, where $\Delta \mu_\text{up}$ is the estimated uncertainty on the upper limit.
\item
  \texttt{--clsAcc}: Determines the absolute accuracy up to which the value of \CLs (or $p_{\mu}$) values are computed when searching for the upper limit.
  The default is 0.5\%.
\item
  \texttt{-T} or \texttt{--toysH}: Determines the minimum number of pseudo-data sets that are generated for each value of $\mu$ with a default value of 500.
\end{itemize}
The distributions of the test statistic for the pseudo-data sets generated assuming each value of $\mu$ that is tested during the bisection algorithm and $\mu=0$ can be saved in the output file by specifying  the option \ttseq{--saveHybridResult}.
Figure~\ref{fig:exampleLHCteststat} shows the  distributions of the test statistic $\widetilde{q}_{\mathrm{LHC}}(\mu=0.4)$ for $\mu=0$ and 0.4 in pseudo-data sets obtained with Datacard~\ref{dc:template}. 

\begin{figure}[htbp]
\centering
\includegraphics[width=\cmsFigWidth]{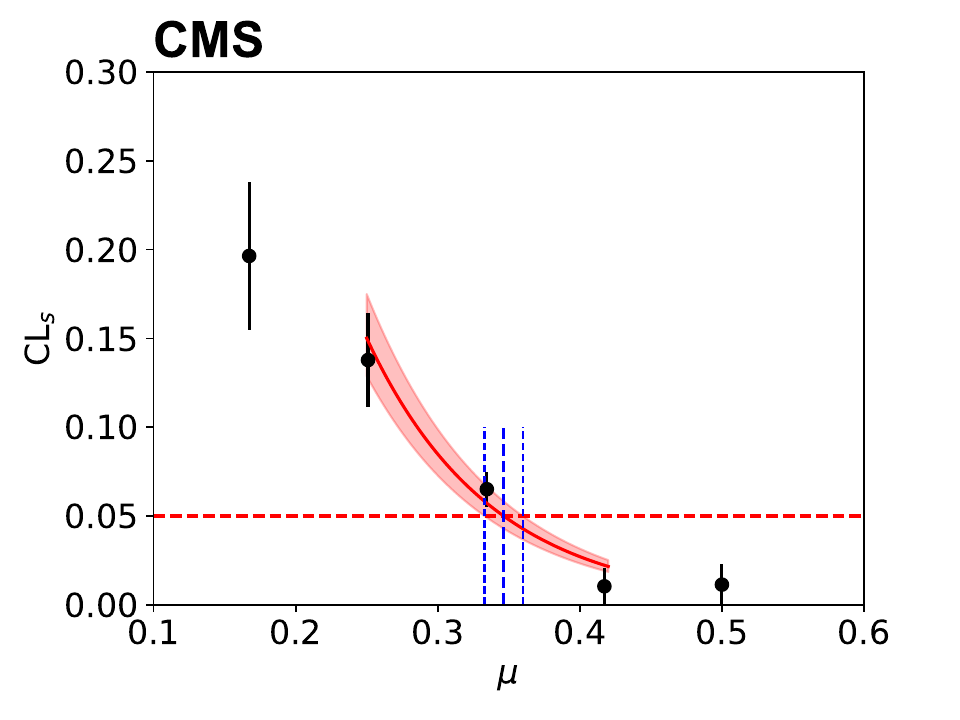}
\caption{
Calculated \CLs as a function of $\mu$, used to determine the 95\% \CL upper limit for Datacard~\ref{dc:template}.
The solid red line is used to interpolate the \CLs values to find the crossing at 0.05, and the shaded band indicates the uncertainty in the interpolation that is used to estimate an uncertainty in the upper limit. 
The vertical dashed blue lines show the derived upper limit and the estimated uncertainty due to the number of pseudo-data sets used in the calculation.}
\label{fig:limitscan}
\end{figure}

To further improve the accuracy when searching for the upper limit, \combine interpolates across several results to estimate $\mu_\text{up}$.
The interpolation uses an exponential function that is fit to the set of results that are closest to the chosen \CL and the range in $\mu$ used for the fit is determined by the accuracy specified in the command line.
A plot of the calculated \CLs value as a function of $\mu$  can be produced using the option \texttt{-{}-plot=name.png}.
Figure~\ref{fig:limitscan} shows the calculation of the upper limit at the 95\% \CL using the \CLs criterion with Datacard~\ref{dc:template}.

The \texttt{AsymptoticLimits} method can be used to calculate the upper limits in statistical models that use the default physics model with the LHC-style prescription. 
This is the default method that will be run if the command line option \ttseq{-M} is not specified. 
In this method, the limit calculation relies on asymptotic approximations for the distributions of the $\widetilde{q}_{\mathrm{LHC}}(\mu)$, following the prescription described in Ref.~\cite{Cowan:2010js}.
The tool also calculates the median and 2.5, 16, 85 and 97.5\% quantiles of the expected distribution of the upper limit assuming $\mu=0$, using an Asimov data set.
The output \ttree contains an entry for each of these results, which can be identified by the \ttseq{quantileExpected} branch.

By default upper limits are calculated at the 95\% \CL ($\alpha=0.05$).
This can be modified using the option \ttseq{-{}-cl=<X>} where \texttt{X} is $(1-\alpha)$.
Upper limits are calculated using the \CLs criterion by default.
Alternatively, it is possible to only use  $p_{\mu}$  by specifying the option \texttt{-{}-rule Pmu} in the command line.
It is also possible to calculate the values of \CLs and $p_{\mu}$ for a single value of $\mu$, bypassing the bisection algorithm, by specifying \ttseq{-{}-singlePoint <r>}, where \texttt{r} is the desired value of $\mu$.

The \ttseq{HybridNew} method can also be used to compute Feldman--Cousins intervals by specifying the option \texttt{--LHCmode} \ttseq{LHC-feldman-cousins}.
This method allows for calculating confidence intervals with accurate coverage both in scenarios with low event counts or where physical boundaries are placed on the parameters of interest $\mu$. 
For example, this method has been used in Higgs boson property measurements at CMS~\cite{CMS:2013fjq}.
The following procedure can be used to produce one-dimensional confidence intervals or multidimensional confidence regions for physics models with multiple parameters of interest:
\begin{itemize}
  \itemsep1pt\parskip0pt\parsep0pt
  \item
    For each parameter point, run the \ttseq{HybridNew} method with the option \ttseq{--LHCmode} \ttseq{LHC-feldman-cousins} \ttseq{--singlePoint}  
    \texttt{\seqsplit{<mu1>=<v1>,<mu2>=<v2>,<mu3>=<v3>,...}} 
    \ttseq{--saveHybridResult} to generate the distributions of the test statistic
    \begin{linenomath*}
    \begin{equation}
    q_{\mathrm{FC}}(\vec{\mu}) = \text -2\ln\left(\dfrac{\Likelihood(\vec{\mu},\hat{\hat{\vtheta}}({\vec{\mu}}))}{\mathcal{L}_{\Omega}(\hat{\vec{\mu}},\hat{\vtheta})}\right),
    \end{equation}
  \end{linenomath*}
    where $\vec{\mu}=\mu_{1},\mu_{2},...$ are the parameters of interest and $\mathcal{L}_{\Omega}(\hat{\vec{\mu}},\hat{\vtheta})$ indicates the maximum of the likelihood function within the bounded region $\Omega$ of its parameters.
    This region can be defined using the command line option \ttseq{--setParameterRanges} 
    \texttt{\seqsplit{<mu1>=<mu1min>,<mu1max>:<mu2>=<mu2min>,<mu2max>:...}} 
    This step also calculates the value of the test statistic for the observed data $q^{\text{obs}}_{\mathrm{FC}}(\vec{\mu})$.
  \item
    Collect the resulting output files into a single \Root file and find the set of points for which
    \begin{linenomath*}
    \begin{equation}
      \int_{q^{\text{obs}}_{\mathrm{FC}}(\vec{\mu})}^{\infty} f(q_{\mathrm{FC}}(\vec{\mu})|\vec{\mu})\,\rd{q_{\mathrm{FC}}} > \alpha,
    \end{equation}
    \end{linenomath*}
    to form the $100 (1-\alpha)\%$ \CL allowed region. 
  \item The output \Root file contains the test statistic value for each pseudo-data set as \Roofit \ttseq{RooStats::HybridResult} objects. These can be used to determine confidence intervals or contours at different values of $\alpha$.
\end{itemize}

\subsection{Significance calculation}
\begin{figure}[htbp]
  \centering
  \includegraphics[width=\cmsFigWidth]{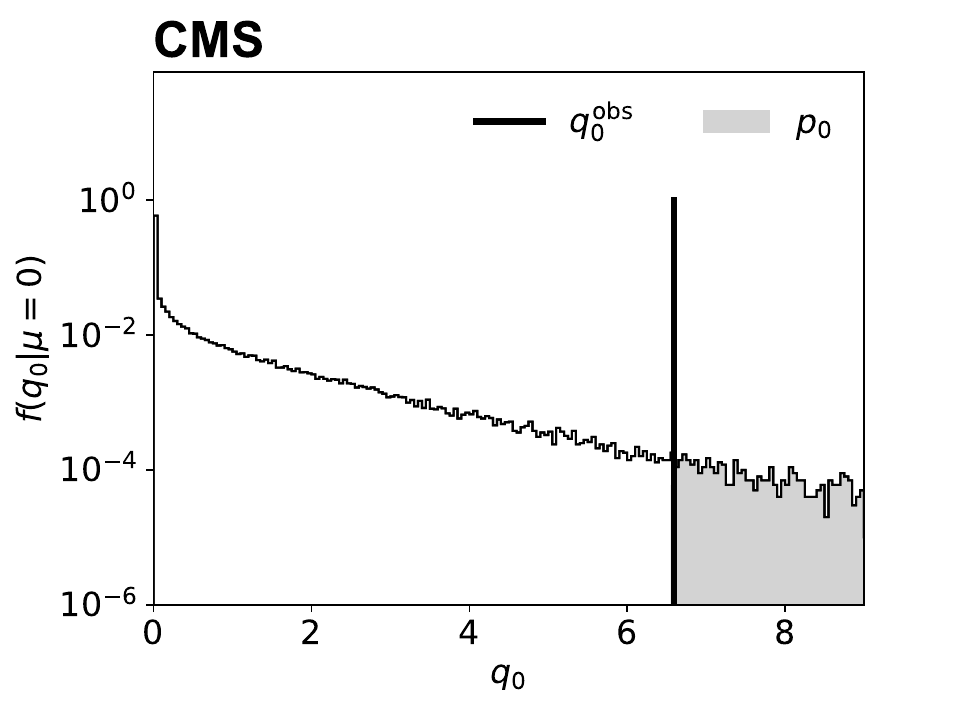}
  \caption{Distribution of $q_{0}$ in 100,000 pseudo-data sets from Datacard~\ref{dc:param}.
  The observed value of the test statistic is indicated by the black vertical line and the region used to determine $p_{0}$ is indicated by the light gray shaded region.}
  \label{fig:exampleSignif}
\end{figure}

The \texttt{HybridNew} method is also used to calculate the significance of the observed data when considered against a null hypothesis. 
For statistical models constructed using the default physics model, this estimates the significance of the presence of a signal contribution in the data where the null hypothesis represents the absence of the signal. 
By specifying the options \ttseq{--LHCmode} \ttseq{LHC-significance}, \combine generates pseudo-data under the background-only hypothesis ($\mu=0$) and evaluates the test statistic $q_{0}$ defined by
\begin{linenomath*}
  \begin{equation}
    q_{0} =  \begin{cases} -2\ln\left(\frac{\Likelihood(0,\hat{\hat{\vtheta}}({0}))}{\Likelihood(\hat{\mu},\hat{\vtheta})}\right) & \text{if } \hat{\mu}>0 \\
      0 & \text{otherwise}, 
      \end{cases}
  \end{equation}
\end{linenomath*}
for each pseudo-data set.
The value of the test statistic for the observed data $q^{\text{obs}}_{0}$ is also calculated in order to determine the $p$-value
\begin{linenomath*}
  \begin{equation}
    p_{0} = \int_{q^{\text{obs}}_{0}}^{\infty} f(q_{0}|0)\,\rd{q_{0}},
  \end{equation}
\end{linenomath*}
where $f(q_{0}|0)$ is the distribution of the test statistic determined using the pseudo-data sets.
Figure~\ref{fig:exampleSignif} shows the observed value of $q_{0}$ and the distribution of $q_{0}$ in pseudo-data sets assuming $\mu=0$ for Datacard~\ref{dc:param}.

The value of $p_{0}$ and corresponding significance of the signal in the example parametric analysis can be calculated using 100,000 pseudo-data sets with Datacard~\ref{dc:param} using the command line below: 
\begin{lstlisting}[style=commandline]
$ combine datacard-3-parametric-analysis.txt -M HybridNew --LHCmode LHC-significance -T 100000 --mass 125
\end{lstlisting}
The results are output to the terminal as: 
\begin{lstlisting}[style=output]
>  -- Hybrid New --
> Significance: 2.54397  -0.0146063/+0.0151701
> Null p-value: 0.00548 +/- 0.000233452
> Done in 5.95 min (cpu), 7.57 min (real)
\end{lstlisting}
The value of $p_{0}$ is converted into a significance using a standard normal distribution~\cite{Workman:2022ynf}. 
The method \ttseq{Significance} can be used in order to speed up the calculation using the asymptotic approximation for the distribution $f(q_{0}|\mu=0)$ given in Ref.~\cite{Cowan:2010js} thereby avoiding the need to generate pseudo-data in cases where the number of events in data is large. For the same datacard, the asymptotic approximation for the significance can be calculated with:
\begin{lstlisting}[style=commandline]
$ combine datacard-3-parametric-analysis.txt -M Significance --mass 125
\end{lstlisting}
The result of the calculation using the asymptotic approximation is output to the terminal as: 
\begin{lstlisting}[style=output]
>  -- Significance --
> Significance: 2.56729
> Done in 0.00 min (cpu), 0.00 min (real)
\end{lstlisting}
The significance result and its estimated uncertainty are stored in the output \Root file in the \ttseq{limit} and \ttseq{limitErr} branches, respectively. 
Using the \ttseq{--pval} option, the $p$-value is stored instead of the significance.  

\subsection{Bayesian upper limits and credible regions}\label{bayesian-limits-and-credible-regions}
Bayesian calculations in \combine are based on the posterior probability $p(\vec{\mu})$ defined by
\begin{linenomath*}
  \begin{equation}
    p(\vec{\mu}) = 
    \frac{1}{p(\left\{\vec{x}\right\}_{d})}\int
    \mathcal{L}\left(\vec{\Phi}\right)\prod_{k}\pi_{k}(\knuisance) \rho(\vec{\mu})\,\rd\vtheta,
  \end{equation}
\end{linenomath*}
where the index $d$ runs over the events in the observed data set $\left\{\vec{x}\right\}_{d}$, and $p(\left\{\vec{x}\right\}_{d})$ is defined such that \ifthenelse{\boolean{cms@external}}{\\}{}$\int p(\vec{\mu})\,\rd\vec{\mu} = 1$. 
The prior term for the parameters of interest $\rho(\vec{\mu})$ must be specified by the user. 
By default this prior is assumed to be uniform over the ranges $(a_{i},b_{i})$ specified for each parameter of interest $\mu_{i}$,
\begin{linenomath*}
\begin{equation}
  \rho(\vec{\mu})\propto\prod_{i}\mathcal{U}(\mu_{i};a_{i},b_{i}).
\end{equation}
\end{linenomath*}

Bayesian upper limits are calculated in \combine using either the \ttseq{BayesianSimple} method for relatively simple statistical models or the \ttseq{MarkovChainMC} method for models with multiple parameters of interest or nuisance parameters, which perform the marginalization over the nuisance parameters.
For statistical models with a single parameter of interest $\mu$, the prior can be modified via the command line using the option \ttseq{--prior} \ttseq{<prior>} with the following options available:
\begin{itemize}
  \item \texttt{flat}: the default uniform prior.
  \item \ttseq{1/sqrt(r)}: inverse square root prior. This is the Jeffreys prior for the mean of a Poisson distribution~\cite{Jeffreys}. 
  \item Any valid \Root \ttseq{TMath::Formula} expression with \texttt{@0} as the parameter of interest.
  \item Any string that names a \RooFit \ttseq{RooAbsPdf} object contained in one of the input workspaces. This option can also be used for statistical models with multiple parameters of interest.
\end{itemize}

Both methods compute the $100 (1-\alpha)\%$ credible upper limit on the parameter of interest $\mu_\text{up}$ as
\begin{linenomath*}
\begin{equation}
  \int_{-\infty}^{\mu_\text{up}} p(\mu)\,\rd\mu = 1-\alpha.
\end{equation}
\end{linenomath*}
The value of $\alpha$ can be modified by specifying the option \ttseq{--cl=}$1-\alpha$.

The \ttseq{BayesianSimple} method computes $\mu_\text{up}$ using numerical integration, while the 
\texttt{\seqsplit{MarkovChainMC}} 
method uses Markov chain integration~\cite{Moneta:2010pm}.

The number of steps in the Markov chain and number of  chains to compute can be specified via the command line, as well as the number of steps to ignore from the start of the chain.
The user can specify a proposal algorithm by which the Markov chain evolves using the option \ttseq{--proposal} \ttseq{<algorithm>} with the following options:
\begin{itemize}
  \itemsep1pt\parskip0pt\parsep0pt
  \item
    \texttt{uniform}: Selects the next parameter point in the chain at random.
  \item
    \texttt{gaus}: Uses a product of independent normal distributions, one for each nuisance parameter where the standard deviation of the distribution for each variable is set to some fraction of the range of the parameter defined by the option 
    \texttt{\seqsplit{--propHelperWidthRangeDivisor}}. 
  \item
    \texttt{ortho}: This is the default proposal and is similar to the \texttt{gaus} proposal except that at each point in the chain, only a single parameter is varied.
  \item
    \texttt{fit}: With this proposal, \combine computes the Hessian matrix of $-2\ln p(\vec{x};\mu,\vtheta)$ with respect to the nuisance parameters $\vtheta$ to construct the proposal function. The accuracy can be improved by including the option \ttseq{--runMinos}.
\end{itemize}

The value of  $\mu_\text{up}$ and an estimate of its uncertainty are obtained from the average over $N$ independent Markov chains, specified by the command line option \texttt{--tries <N>}. 
The 95\% Bayesian upper limit on the default physics model parameter $r$ for Datacard~\ref{dc:counting} using 100 Markov chains can be calculated using the following:
\begin{lstlisting}[style=commandline]
$ combine datacard-1-counting-experiment.txt -M MarkovChainMC --tries 100
\end{lstlisting}
These values are saved in the output \Root file and output to the terminal as below: 
\begin{lstlisting}[style=output]
>  -- MarkovChainMC --
> Limit: r < 2.21031 +/- 0.0133576 @ 95% credibility (100 tries)
> Done in 0.05 min (cpu), 0.05 min (real)
\end{lstlisting}

If using the \ttseq{MarkovChainMC} method, it is also possible to store the resulting Markov chains in the output file from \combine using the option \ttseq{--saveChain}. 
This allows for estimating the posterior distribution $p(\vec{\mu})$ for one or more parameters of the model and deriving credible intervals or regions.

\subsection{Maximum likelihood estimates and scans}\label{likelihood-fits-and-scans}
Likelihood-based parameter estimation is performed in \combine using the \texttt{MultiDimFit} method. This method can be used to calculate maximum likelihood estimates for the parameter values and their uncertainties through several different approaches. 
This method has been used by the CMS Collaboration to provide measurements in a number of different scenarios, including Higgs boson production and decay rates, and measurements of the Higgs boson couplings~\cite{CMS:2022dwd}.
The \texttt{MultiDimFit} method is used to evaluate the negative-log-likelihood function, ${-}\ln\Likelihood(\vec{\Phi})$, and obtain maximum likelihood estimates and estimates of confidence intervals for the parameters of interest $\vec{\mu}$ using the profile likelihood ratio,
\begin{linenomath*}
\begin{equation}\label{eqn:qmu}
  q(\vec{\mu}) = -\ln\left(\frac{\Likelihood(\vec{\mu},\hat{\hat{\vtheta}}({\vec{\mu}}))}{\Likelihood(\hat{\vec{\mu}},\hat{\vtheta})}\right),
\end{equation}
\end{linenomath*}
where $\hat{\vec{\mu}}$ are the maximum likelihood estimates for the parameters of interest, and $\hat{\hat{\vtheta}}({\vec{\mu}})$ and $\hat{\vtheta}$ are the values of the nuisance parameters for which $\mathcal{L}$ is maximized for a specific set of parameter values $\vec{\mu}$ and for the maximum likelihood estimates $\hat{\vec{\mu}}$, respectively.
The process of finding the parameter values that maximize the likelihood function is typically referred to as a ``fit''. 
In Eq.~(\ref{eqn:qmu}) there are two such fits. The one in the denominator is often referred to as the ``overall best fit''. 
The default parameter values are commonly referred to as ``pre-fit'', while the maximum likelihood estimates are commonly referred to as ``post-fit''. 
Throughout this and following sections, the process of determining the maximum likelihood estimates is referred to as maximum likelihood optimization so as not to confuse this procedure with the more general ``goodness of fit'' methods described in Section~\ref{goodness-of-fit-tests}.

The following choices for the \ttseq{--algo} option are supported in \combine:
\begin{itemize}
\item
  \texttt{none}: this is the default algorithm.
  The algorithm finds the parameter values $\hat{\vec{\Phi}}$ that maximize $\Likelihood(\vec{\Phi})$  and reports the maximum likelihood estimates of the parameters of interest.
  For a model with $N$ parameters of interest, the output \ttree contains $N$ branches, one for each parameter of interest with the maximum likelihood estimates. The output of this algorithm can be used as the starting point for the other algorithms to reduce their evaluation time.
\item
  \texttt{singles}: the algorithm determines the maximum likelihood estimates for each parameter of interest $\mu$ and sequentially determines 68\% confidence intervals for each parameter of interest. 
  The output \ttree contains one branch for each parameter of interest.
  One of the entries contains the maximum likelihood estimates with \texttt{quantileExpected} set to $-1$. Two additional entries for each parameter of interest provide the upper $\mu^{+}$ and lower $\mu^{-}$ bounds of the 68\% \CL interval for that parameter determined as the range of that parameter for which $q(\mu)< 1/2$.
\item
  \texttt{cross}: the algorithm determines the maximum likelihood estimates for each parameter of interest and a set of intervals for each parameter $\mu_{l}$ for which $q(\mu_{l})< \frac{c}{2}$, where $c$ is determined by $\int_{c}^{\infty}\chi^{2}(x;n)\,\rd{x} = \alpha$ and $\chi^{2}(x;n)$ is a chi-squared distribution with the number of degrees of freedom equal to the number of parameters of interest $n$.
  The output \ttree has one entry with the maximum likelihood estimates for the parameters of interest $\hat{\mu}$, and two entries for each parameter of interest, corresponding to the points that define the interval.
\item
  \texttt{contour2d}: for statistical models with two parameters of interest, this algorithm constructs two dimensional contours bounding the regions for which $q(\vec{\mu})< \frac{c}{2}$, where $c$ is computed from $\int_{c}^{\infty}\chi^{2}(x;2)\,\rd{x} = \alpha$. The output  contains values corresponding to the maximum likelihood estimates with \ttseq{quantileExpected} set to $-1$, and additional entries that define the contour. For $\alpha=0.32$ the contour produced corresponds to $c = 2.3$~\cite{Workman:2022ynf}.  
\item
  \texttt{random}: in this algorithm, the value of $q(\vec{\mu})$ is evaluated for $N$ uniformly distributed random points in the space spanned by the parameters of interest. The number of points is set by the option \ttseq{--points=<N>}.
\item
  \texttt{fixed}: in this algorithm, the value of $q(\vec{\mu})$ is calculated at a specified point in the space spanned by the parameters of interest. The fixed point is specified using the \ttseq{--fixedPointPOIs} option.
\item
  \texttt{grid}: the value of $q(\vec{\mu})$ is evaluated on a grid of parameter of interest points with $N$ points in total. The number of points $N$ is specified with the option \ttseq{--points=<N>}.
  It is possible to partition the scan into multiple runs of \Combine using the options \ttseq{--firstPoint} \texttt{<n>} and \ttseq{--lastPoint} \texttt{<m>}, where ${0}\leq$\texttt{n}$<$\texttt{m}$\leq N$.
\end{itemize}
The \texttt{fixed} and \texttt{grid} algorithms can be used to evaluate the likelihood function and $q(\vec{\mu})$ for any value of the statistical model parameters. 
In each of the above options the value of $\alpha$ can be defined using the option \ttseq{--cl=}$1-\alpha$.
The output \ttree branch named \ttseq{deltaNLL} stores the value of $q(\vec{\mu})$. 
Additional branches can be included to store the values of any nuisance parameter, any \ttseq{RooCategory} object, or any \texttt{RooFormulaVar} object contained in the workspace produced by \Combine by adding the options \ttseq{--saveSpecifiedNuis}, 
\texttt{\seqsplit{--saveSpecifiedIndex}}, 
or \ttseq{--saveSpecifiedFunc}, respectively.

Confidence intervals calculated with \combine using these algorithms will yield results with   approximately correct coverage in the absence of large non-Gaussian uncertainties~\cite{Wilks, Wald,Engle}. 
If the overall best fit values lie on physical  boundaries in the parameter space, the  Feldman--Cousins procedure described in Section~\ref{sec:intervals} should be used to obtain intervals with improved coverage properties.

\begin{figure}[htbp]
  \centering
  \includegraphics[width=\cmsFigWidth]{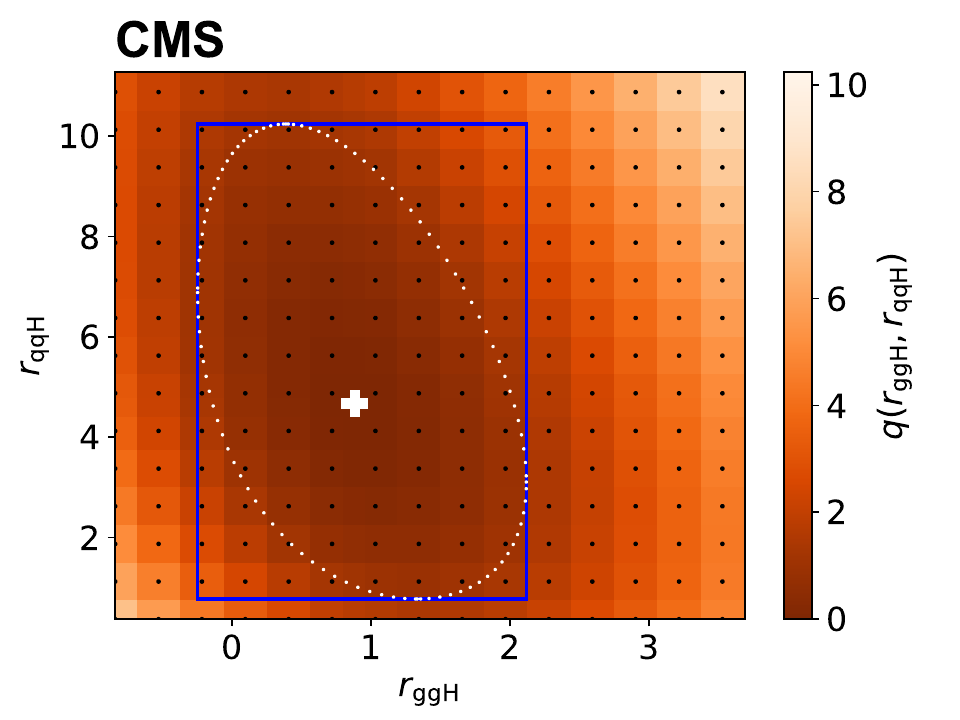}
  \caption{
    Values of $q(\rggH,\rqqH)$ for Datacard~\ref{dc:multisig} in a model with two parameters of interest $\rggH$ and $\rqqH$.
    The orange scale shows the values obtained in \combine at the set of points indicated by the black dots, using the \texttt{grid} algorithm. The blue box is constructed using the \texttt{cross} algorithm with $(1-\alpha)=0.68$.
    The white cross and white dots indicate, respectively, the maximum likelihood estimates for $\rggH$ and $\rqqH$ from the best fit, and the 68\% \CL confidence region obtained using the \texttt{contour2d} algorithm defined as the values of $(\rggH,\rqqH)$ for which $q(\rggH,\rqqH)= 2.3$.}
  \label{fig:examplescan}
\end{figure}

The maximum likelihood estimates for the parameters $\rggH$ and $\rqqH$ and estimates of their uncertainties can be obtained using Datacard~\ref{dc:multisig} with the \ttseq{PhysicsModel:floatingXSHiggs} physics model using the commands below,
\begin{lstlisting}[style=commandline]
$ text2workspace.py datacard-5-multi-signal.txt -P HiggsAnalysis.CombinedLimit.PhysicsModel:floatingXSHiggs --PO modes=ggH,qqH -o datacard-5-multi-signal.root --mass 125
$ combine datacard-5-multi-signal.root -M MultiDimFit --algo singles --mass 125
\end{lstlisting}
The output from \combine is given below: 
\begin{lstlisting}[style=output]
>  --- MultiDimFit ---
> best fit parameter values and profile-likelihood uncertainties:
>    r_ggH :    +0.882   -0.749/+0.795 (68%)
>    r_qqH :    +4.683   -2.746/+3.464 (68%)
> Done in 0.00 min (cpu), 0.04 min (real)
\end{lstlisting}
For each parameter, the result is presented as a measurement $\hat{\mu}~-\Delta^{-}\mu/+\Delta^{+}\mu$, where the values of $\Delta^{\pm}\mu$ are determined from the 68\% confidence intervals as
\begin{linenomath*}
\begin{equation}
  \Delta^{\pm}\mu = \abs{\mu^{\pm} -\hat{\mu}}.
\end{equation}
\end{linenomath*}
The relative precision on $\rqqH$ is better than for  $\rggH$ due to the larger signal to background ratio in the \texttt{dijet} channel of this datacard. 
The parameters are correlated due to the contribution of each process across both channels. This can be seen by considering the shape of the function $q(\rggH,\rqqH)$.
Figure~\ref{fig:examplescan} shows $q(\rggH,\rqqH)$ using the same datacard and physics model. 
The points indicate the output from \combine using the \texttt{grid} and \texttt{contour2d} algorithms.
The box shown in the figure is constructed from the set of intervals calculated using the \texttt{cross} algorithm with $(1-\alpha)=0.68$. 

\begin{figure}[htbp]
  \centering
  \includegraphics[width=\cmsFigWidth]{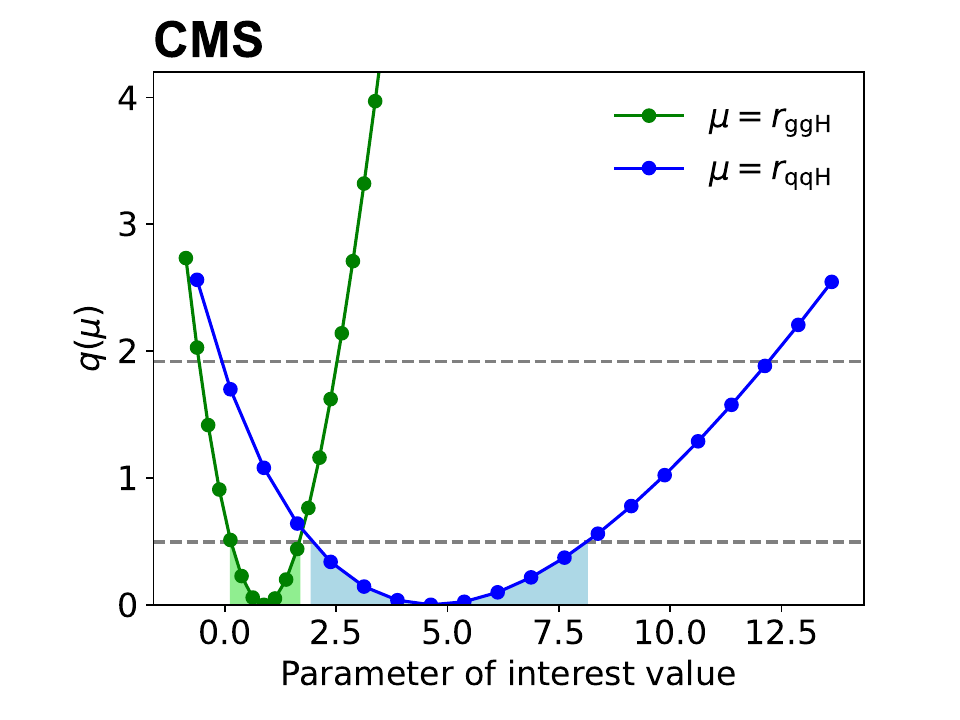}
  \caption{
    Example of $q(\rggH,\hat{r}_{\PQq\PQq\PH})$ and $q(\rqqH,\hat{r}_{\Pg\Pg\PH})$ obtained from \combine with Datacard~\ref{dc:multisig}.
    The points indicate the values at which the functions are evaluated using the \texttt{grid} algorithm, and the shaded region indicates the 68\% \CL intervals on each parameter obtained using the \texttt{singles} algorithm. The horizontal dashed lines indicate the values of $q(\mu)$ used to define 68\% and 95\% \CL intervals. }
  \label{fig:examplescan1D}
\end{figure}

A subset of parameters for any of the algorithms can be specified using multiple instances of the option \ttseq{--parameters}.
In this case, all other parameters of interest are set equal to their default values throughout the calculation.
This behavior can be adjusted by including the option \ttseq{--floatOtherPOI=1}, which instructs \Combine to include the remaining parameters of interest in the list of profiled parameters for the purposes of the calculation of profile likelihoods.
This is not the same as using the \ttseq{--redefineSignalPOIs} option as selecting subsets of parameters does not remove the associated probability distribution $p(\ttheta;\anuisance)$ when nuisance parameters are included. 
Figure~\ref{fig:examplescan1D} shows the values of $q(\rggH,\hat{r}_{\PQq\PQq\PH})$ and $q(\rqqH,\hat{r}_{\Pg\Pg\PH})$ obtained from \combine using the \texttt{grid} algorithm, and the 68\% \CL intervals on each parameter obtained using the \texttt{singles} algorithm. 
In this case, the options \ttseq{--parameters=<X>} \ttseq{--floatOtherPOI=1} are included where \texttt{X} is either \texttt{r\_ggH} or \texttt{r\_qqH}. 

\subsection{Goodness of fit tests and diagnostics}\label{goodness-of-fit-tests}
The \ttseq{GoodnessOfFit} method can be used to perform  a test of compatibility between the observed data and the statistical model.
While the statistical model constructed by \Combine is used as the null distribution,  goodness of fit (GoF) tests do not have a well-defined alternative hypothesis, unlike conventional hypothesis tests.

The \ttseq{GoodnessOfFit} method evaluates a test statistic for the observed data and generates the expected distribution of the test statistic under the null distribution using pseudo-data. 
In \combine this is done in two separate commands: 
\begin{lstlisting}[style=commandline]
$ combine -M GoodnessOfFit <datacard.txt> --algo=<test-statistic>

$ combine -M GoodnessOfFit <datacard.txt> --algo=<test-statistic> --toys <N> 
\end{lstlisting}
The first command evaluates the observed test statistic and the second determines the distribution of the test statistic under the null hypothesis using pseudo-data sets.
The statistical model parameters used to specify the null hypothesis can be set using the \ttseq{--setParameters} option.

All of the supported test statistics are based on binned data. For statistical models that include parametric shape analyses, an automatic binning is applied to the data to evaluate the probability density and calculate the test statistic, unless a specific binning is specified for the observables using the \ttseq{RooAbsReal::setBinning} method. 
If no binning is specified, 100 uniform bins will be used for each observable.  
The following test statistics for the \ttseq{--algo} option are supported in \combine:
\begin{itemize}
\item
  {saturated}:  evaluates a GoF test statistic defined as
  \begin{linenomath*}
  \begin{equation}
    t = -2\ln\frac{\Likelihood(\hat{\vec{\Phi}})}{\Likelihood_{S}},
  \end{equation}
\end{linenomath*}
  where $\Likelihood_{S}$ represents the likelihood for the saturated model~\cite{saturated}.
\item
  {KS}:  evaluates
  the Kolmogorov--Smirnov test statistic~\cite{kolmogorov,smirnov1948,James:1019859}, based on the largest
  difference between the cumulative distribution function and the empirical distribution function across all bins in all channels.
\item
  {AD}:   evaluates the
  Anderson--Darling test statistic based on the integral of the difference between the cumulative distribution function and the empirical distribution function across all bins in all channels. This test statistic gives more importance to the tails of the distribution in data~\cite{anderson1952,KSAD}.
\end{itemize}

\begin{figure}[htbp]
  \centering
  \includegraphics[width=\cmsFigWidth]{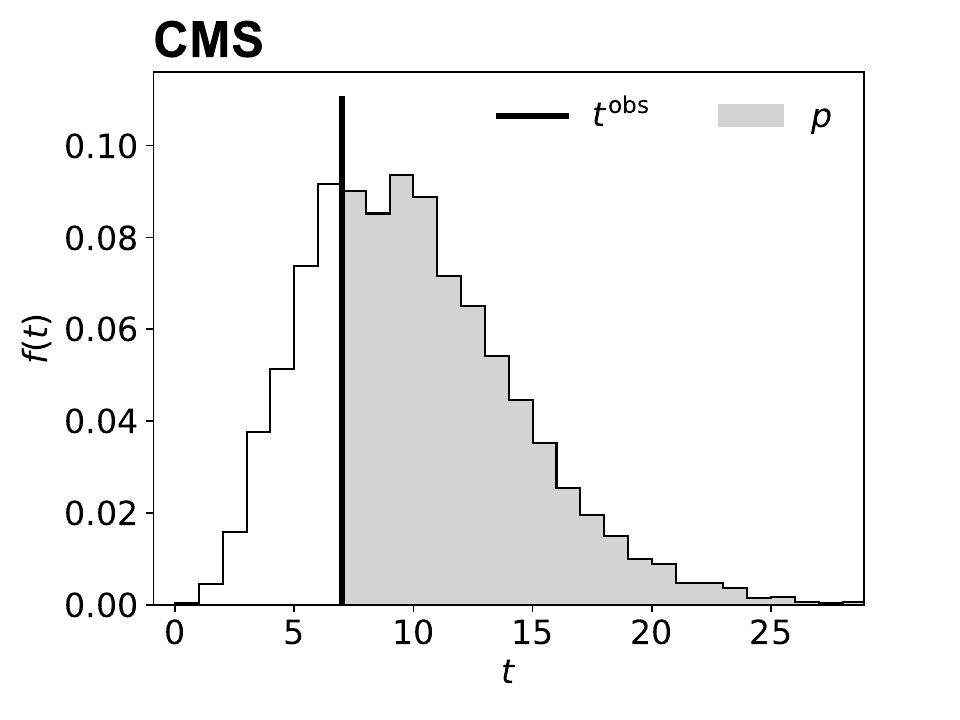}
  \caption{
    Distribution of the saturated test statistic $t$ in 10,000 pseudo-data sets using Datacard~\ref{dc:template}. 
    The observed value of the test statistic is indicated by the black vertical line and the region used to determine $p$ is indicated by the light gray shaded region.}
  \label{fig:examplegof}
\end{figure}

The output \ttree contains a branch called \ttseq{limit} that contains the value of the test statistic in each pseudo-data set if running with the option \ttseq{--toys}, or from the observed data.
These can be used to determine a $p$-value under the null hypothesis.
Figure~\ref{fig:examplegof} shows the distribution of the saturated test statistic $t$ in  pseudo-data sets using the statistical model constructed by \combine from Datacard~\ref{dc:template} with default nuisance parameter values. 
The  $p$-value is 0.73 and is computed from the distribution and the observed value of $t$.  
In this example, the template analysis in  Datacard~\ref{dc:template} was used to generate pseudo-data sets and calculate the corresponding $p$-value. 
The distribution of the test statistic has a peak close to the number of bins for the observable in the datacard, which is expected for this test statistic.  
The method of pseudo-data generation can be modified using the options described in Section~\ref{toy-data-generation}.

The \ttseq{ChannelCompatibilityCheck} method is used to evaluate the compatibility between measurements of the signal rate in the $N_{c}$ separate channels defined in the statistical model.
The method can be used with the default physics model and calculates the value of
\begin{linenomath*}
\begin{equation}
 q = -2\ln\left(\frac{\Likelihood(\hat{r})}{\Likelihood(\hat{r}_{1},\hat{r}_{2},...,\hat{r}_{N_{c}})}\right),
\end{equation}
\end{linenomath*}
where each $r_{i}$, for $i=1,2,\ldots,N$, multiplies the rate of all signal processes in a specific channel $i$, and $\hat{r}$ denotes the maximum likelihood estimate.

The channel compatibility of the multi-signal Higgs boson analysis from Datacard~\ref{dc:multisig} can be calculated using the command line below:  
\begin{lstlisting}[style=commandline]
$ combine datacard-5-multi-signal.txt -M ChannelCompatibilityCheck --mass 125
\end{lstlisting}
The output from these commands is given below: 
\begin{lstlisting}[style=output]
>  --- ChannelCompatibilityCheck ---
> Nominal fit  : r =  1.4812  -0.5987/+0.6644
> Alternate fit: r =  3.5554  -1.8576/+2.3619   in channel dijet
> Alternate fit: r =  1.1506  -0.6393/+0.6939   in channel incl
> Chi2-like compatibility variable: 1.5417
> Done in 0.00 min (cpu), 0.02 min (real)
\end{lstlisting}
These results are not expected to match those from the maximum likelihood estimates for the parameters $\rggH$ and $\rqqH$ as the parameters in that model are related to the process signal multipliers instead of multipliers for the total signal in each channel. 
The compatibility variable in the output is the value of $q$. 
Assuming the values of $\hat{r_{i}}$ are normally distributed, this quantity can be converted to a $p$-value $p=\int_{q}^{\infty}\chi^{2}(x;N_{c}-1)\,\rd{x}$ where $\chi^{2}$ is a chi-squared distribution with $N_{c}-1$ degrees of freedom. 

The output \ttree contains the value of $q$ in the \ttseq{limit} branch.
If the option 
\texttt{\seqsplit{--saveFitResult}} 
is specified, the output \Root file also contains two \ttseq{RooFitResult} objects \ttseq{fit\_nominal}, and \ttseq{fit\_alternate} with the results of the two maximum likelihood optimizations used to calculate the numerator and denominator in $q$, respectively.

The \ttseq{FitDiagnostics}  method provides a number of diagnostic routines to investigate the statistical model.
The method can be used with the default physics model and performs two optimizations of the parameters $\vec{\Phi}$ that maximize the function $\Likelihood(\vec{\Phi})$, the first with the parameter of interest $r$ allowed to vary and the second with $r$ set at a constant value of zero.
The output \ttree contains the maximum likelihood estimator for $r$ and an estimate of its uncertainty.
In addition to the usual output file, an additional \Root file is produced with the name \ttseq{fitDiagnostics}\texttt{\$}\ttseq{NAME.root}.
This file contains additional details about the optimizations performed that can be used to investigate the statistical model and the optimization procedure, the details of which are given in Table~\ref{tab:fitdiag}. 

\begin{table*}[htbp]
  \topcaption{Additional output file contents from the \texttt{FitDiagnostics} method.}
  \centering
  \renewcommand{\arraystretch}{1.6}
  \cmsTable{
    \begin{tabularx}{\textwidth}{lX}
      {Object} & {Description}\\
      \hline
      \texttt{nuisances\_prefit} & \texttt{RooArgSet} containing the default values of $\vtheta$, and their uncertainties estimated from their probability distributions. \\
      \texttt{fit\_b} & \texttt{RooFitResult} object containing the outcome of the maximum likelihood optimization with $r$ fixed to zero. \\
      \texttt{fit\_s} & \texttt{RooFitResult} object containing the outcome of the maximum likelihood optimization with $r$ allowed to vary. \\
      \texttt{tree\_prefit} & \ttree of default values of $\vtheta$ and $\tvtheta$. \\
      \texttt{tree\_fit\_b} & \ttree of maximum likelihood estimates $\hat{\hat{{\vtheta}}}(0)$ and values of $\tvtheta$ from the maximum likelihood optimization fixing $r$ to zero. \\
      \texttt{tree\_fit\_sb} & \ttree of maximum likelihood estimates $\hat{\vtheta}$ and values of $\tvtheta$ from the maximum likelihood optimization allowing $r$ to vary. \\ [\cmsTabSkip]
      \multicolumn{2}{l}{Objects below are present only if the option \texttt{{--plots}} is included.}\\ 
      \texttt{covariance\_fit\_b} & \texttt{TH2D} covariance matrix of the parameters from the maximum likelihood optimization fixing $r$ to zero. \\
      \texttt{covariance\_fit\_s} & \texttt{TH2D} covariance matrix of the parameters from the maximum likelihood optimization allowing $r$ to vary. \\
      \texttt{channel\_observable\_prefit} & \texttt{RooPlot} plot of the probability density, normalized to the yield in data, projected onto each observable in each channel along with a histogram of the data. The statistical model parameters are set to their default values. \\
      \texttt{channel\_observable\_fit\_b} & Same as \texttt{channel\_observable\_prefit}, except the statistical model parameters are set to their maximum likelihood estimates from the optimization fixing $r$ to zero. \\
      \texttt{channel\_observable\_fit\_s} & Same as \texttt{channel\_observable\_prefit}, except the statistical model parameters are set to their maximum likelihood estimates from the optimization allowing $r$ to vary. \\
      \end{tabularx}
  }
  \label{tab:fitdiag}
\end{table*}
In addition to maximum likelihood estimates for all of the model parameters, these results contain estimates of their uncertainties. 
By default, the uncertainties for all nuisance parameters $\Delta\anuisance$ are estimated as $\Delta\knuisance = H^{-1}_{kk}$, where $H_{ij}$ is the Hessian matrix of $q(\hat{\vec{{\mu}}})$, 
\begin{equation}
H_{ij} = \dfrac{\partial^{2} q}{\partial \anuisance_{i}\partial\anuisance_{j}}\bigg\rvert_{\vec{\Phi}=\hat{\vec{\Phi}}},
\end{equation}
where $\hat{\vec{\Phi}}$ are the maximum likelihood estimates of the statistical model parameters. 
The accuracy can be improved by including the option \ttseq{--minos} \ttseq{all}, which determines the uncertainty for each nuisance parameter as 
\begin{linenomath*}
\begin{equation}
  \Delta^{\pm}\knuisance = \abs{\anuisance^{\pm} -\hat{\anuisance}}, 
\end{equation}
\end{linenomath*}
where the range $\left(\anuisance^{-},\anuisance^{+}\right)$ is determined as the region for which $q(\anuisance)< 1/2$, similar to the \texttt{singles} algorithm described in Section~\ref{likelihood-fits-and-scans}. 

The results contained in this file can be used to study the additional constraints imposed by the observed data by considering the difference between the maximum likelihood estimates of the nuisance parameters (``post-fit'') and their uncertainties compared to their default (``pre-fit'') values. 

The \ttseq{FitDiagnostics} method can also be used to calculate nuisance parameter impacts~\cite{ATLAS:2014vuz} for statistical models with one or more parameters of interest, defined with respect to any particular parameter of interest $\mu$ for each nuisance parameter $\knuisance$ as   
\begin{linenomath*}
\begin{equation}
  \Delta\mu^{\pm} = \hat{\mu}\left(\hat{\anuisance}_{k}\pm\Delta^{\pm}\knuisance\right) -\hat{\mu},
\end{equation}
\end{linenomath*}
where $\hat{\mu}(\knuisance\pm\Delta^{\pm}_{k})$ is the value of $\mu$ that maximizes the likelihood function when the nuisance parameter is shifted from its maximum likelihood estimator value by its uncertainty. 
Large impacts typically result from nuisance parameters that contribute significantly to the total uncertainty in $\mu$ and the sign of the impact indicates the sign of the (anti-)correlation between the parameter of interest and that nuisance parameter.
These impacts provide  diagnostic information that encapsulates both the constraints imposed on the nuisance parameter by the data and the correlation of that nuisance parameter with a particular parameter of interest. 
Figure~\ref{fig:tttt} shows the impacts for the default physics model parameter $r$ calculated in the statistical analysis leading to the observation of four top quark production by the CMS Collaboration~\cite{CMS:2023ftu}. 
These impacts are calculated using both the observed data set (``obs.'') and an Asimov data set constructed assuming standard model production of four top quarks to obtain the expected (``exp.'') impacts.  

\ifthenelse{\boolean{cms@external}} { 
\begin{figure*}[htbp!]
}{
\begin{figure}[htbp]
}
  \centering
  \ifthenelse{\boolean{cms@external}} { 
  \includegraphics[width=1.6\cmsFigWidth]{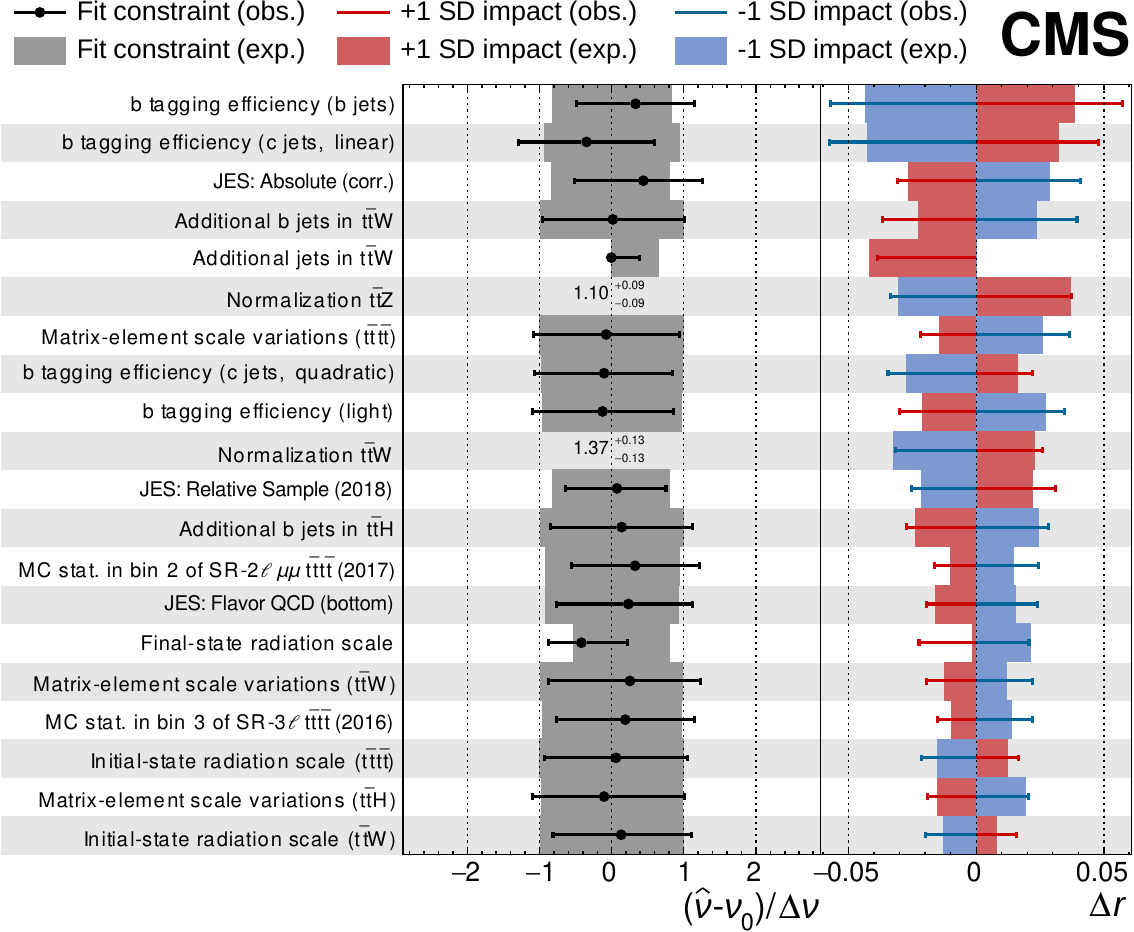}
  }{
  \includegraphics[width=\textwidth]{Figure_009.pdf} 
  }
  \caption{
    Example of nuisance parameter uncertainties and impacts calculated in \combine for the observation of four top quark production. Each row gives the name of the nuisance parameter, the difference in its maximum likelihood estimate $\hat{\anuisance}$ with respect to its default value $\anuisance_{0}$ relative to its uncertainty $\Delta\anuisance$, and the impact with respect to the default physics model parameter $\Delta r$.  The nuisance parameter constraints and impacts are calculated using the observed data set (obs.) and an Asimov dataset constructed assuming standard model production of four top quarks (exp.). The red and blue lines in each row represent the positive impact $\Delta r^{+}$ and negative impact $\Delta r^{-}$, respectively, for the observed data. Similarly, the red and blue shaded boxes represent the same quantities for the Asimov dataset. The error bars on the fit constraint values indicate the ratio of $\Delta^{-}\anuisance$ or $\Delta^{+}\anuisance$, to their default values. The two numerical values displayed in the figure give the value of $\hat{\anuisance}^{+ \Delta^{+}\anuisance}_{- \Delta^{-}\anuisance}$ for two rate parameters, which do not have well-defined default uncertainty values. Figure adapted from Ref.~\cite{CMS:2023ftu}. 
    }
  \label{fig:tttt}
\ifthenelse{\boolean{cms@external}} { 
\end{figure*}
}{
\end{figure}
}

\begin{figure}[htbp!]
  \centering
  \includegraphics[width=\cmsFigWidth]{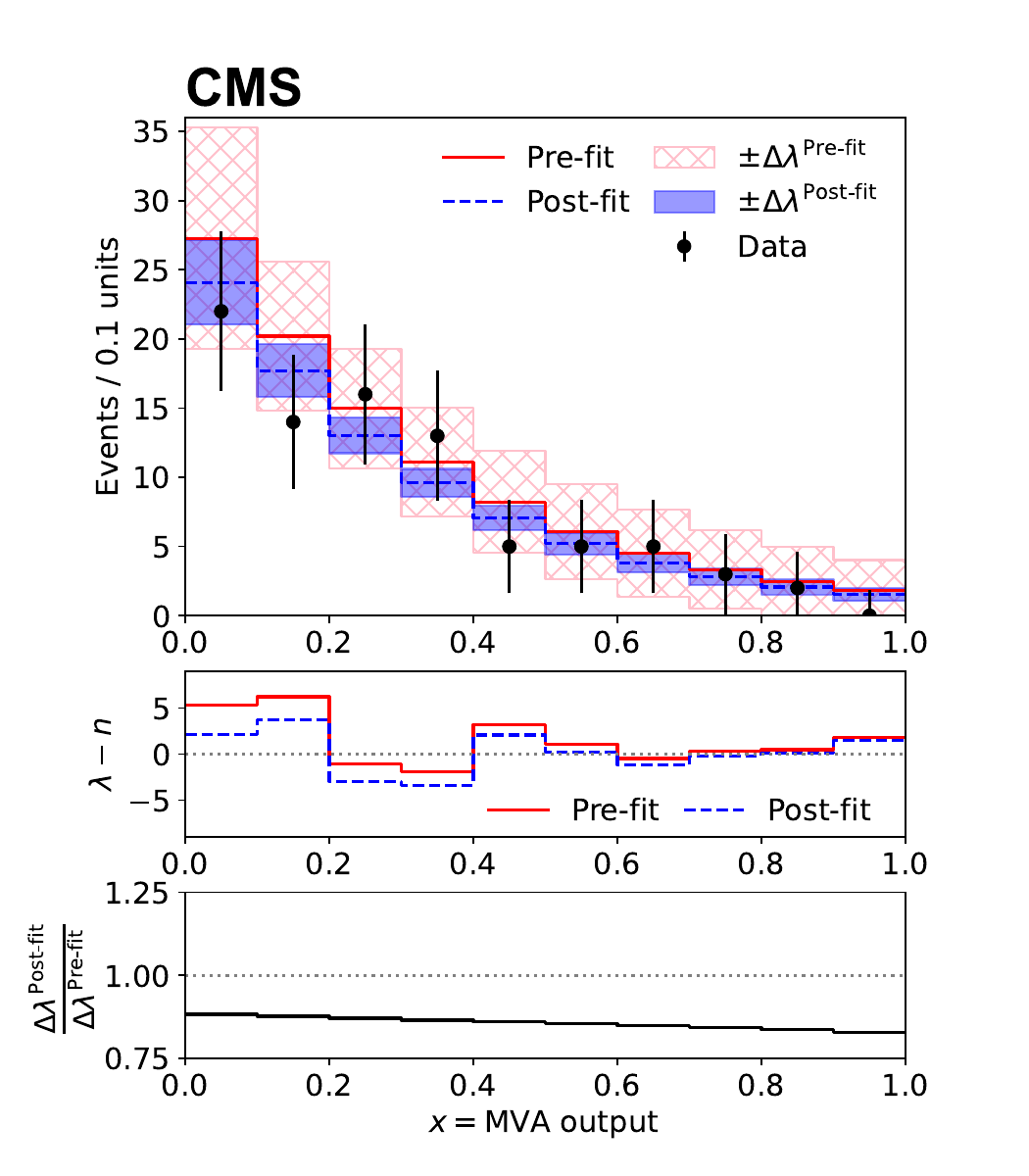}
  \caption{
    Distributions of the observable $x$ for the data and background process in Datacard~\ref{dc:template} and their uncertainties. The upper panel shows the distribution for the default values of the nuisance parameters (red solid line, pre-fit) and for the maximum likelihood estimates assuming no signal (blue dashed line, post-fit). The pink hatched and blue shaded bands show the estimate of the uncertainty in each bin for the pre-fit and post-fit distributions, respectively. The middle panel shows the difference between the expected number of events in the background processes  ($\lambda$) and the data ($n$) in the pre-fit (red solid line) and post-fit (blue dashed line) cases, and the lower panel shows the ratios of the estimated uncertainties of the post-fit distribution $\Delta\lambda^{\text{Post-fit}}$ to the pre-fit $\Delta\lambda^{\text{Pre-fit}}$ in each bin. 
    }
  \label{fig:postpre}
\end{figure}

By including the option \ttseq{--saveShapes} with this method, \combine  saves  additional \Root \ttseq{TH1F} histogram objects in the output file.
These represent the contributions from each process in each channel of the statistical model evaluated at the default values of the parameters, and at the maximum likelihood estimates from the two optimizations. 
This allows a visualization of the pdfs of the primary observables at the pre-fit and post-fit values of the statistical model parameters individually for each process and each channel.  
For statistical models that include parametric shape analyses, an automatic binning is used to construct the histograms unless a particular binning is specified for the observables using the \ttseq{RooAbsReal::setBinning} method. 
If no binning is specified, 100 uniform bins will be used for each observable.
The total signal, total background, and total contributions in each channel are also saved as separate \ttseq{TH1F} histogram objects.
By adding the option \ttseq{--saveWithUncertainties}, the output  also includes estimates of the covariance between each of the bin yields, accounting for any correlations between the parameters of the statistical model.
This is achieved using the \texttt{RooFitResult::randomizePars()} method from the results of the optimization.
Figure~\ref{fig:postpre} shows the distribution of the observable $x$ for the data and the background process in Datacard~\ref{dc:template} using the default physics model. 
The uncertainties are estimated by sampling from the distributions $p(\anuisance|\ttheta)$ in the pre-fit case and from the covariance matrix of the model parameters obtained from a fit to the data shown assuming $r=0$ (post-fit). 
The change in the expected number of events in each bin $\lambda$ and their uncertainties $\Delta\lambda$ visually indicate the additional constraint imposed by the data on the statistical model parameters. 

The \ttseq{FitDiagnostics} method can be run with the \ttseq{--toys} option such that the process normalization and bin yields in each channel resulting from each of the two optimizations performed using pseudo-data sets are stored, allowing for more detailed studies of the statistical model.

\section{Summary}\label{sec:summary}
After a decade of development, the \combine package has become the main tool used for statistical analysis of data by the CMS Collaboration.
The tool is based on the \Root~\cite{Brun:1997pa}, \RooFit~\cite{Verkerke:2003ir}, and \RooStats~\cite{Verkerke:2003ir} software packages to provide a command-line interface to several common statistical workflows used in high-energy physics.
The statistical model is constructed from a text file provided by the user and a configurable physics model that encodes the parameters of interest and the nuisance parameters that model systematic uncertainties.
The \combine package can perform a variety of statistical procedures including calculating confidence or credible intervals, evaluating profile likelihoods, and performing goodness of fit tests.
The online documentation~\cite{combinedoc} contains comprehensive information on the capabilities and instructions for running the \combine package, as well as detailed instructions for its installation.

\begin{acknowledgments}
  We congratulate our colleagues in the CERN accelerator departments for the excellent performance of the LHC and thank the technical and administrative staffs at CERN and at other CMS institutes for their contributions to the success of the CMS effort. In addition, we gratefully acknowledge the computing centres and personnel of the Worldwide LHC Computing Grid and other centres for delivering so effectively the computing infrastructure essential to our analyses. Finally, we acknowledge the enduring support for the construction and operation of the LHC, the CMS detector, and the supporting computing infrastructure provided by the following funding agencies: SC (Armenia), BMBWF and FWF (Austria); FNRS and FWO (Belgium); CNPq, CAPES, FAPERJ, FAPERGS, and FAPESP (Brazil); MES and BNSF (Bulgaria); CERN; CAS, MoST, and NSFC (China); MINCIENCIAS (Colombia); MSES and CSF (Croatia); RIF (Cyprus); SENESCYT (Ecuador); ERC PRG, RVTT3 and MoER TK202 (Estonia); Academy of Finland, MEC, and HIP (Finland); CEA and CNRS/IN2P3 (France); SRNSF (Georgia); BMBF, DFG, and HGF (Germany); GSRI (Greece); NKFIH (Hungary); DAE and DST (India); IPM (Iran); SFI (Ireland); INFN (Italy); MSIP and NRF (Republic of Korea); MES (Latvia); LMTLT (Lithuania); MOE and UM (Malaysia); BUAP, CINVESTAV, CONACYT, LNS, SEP, and UASLP-FAI (Mexico); MOS (Montenegro); MBIE (New Zealand); PAEC (Pakistan); MES and NSC (Poland); FCT (Portugal); MESTD (Serbia); MCIN/AEI and PCTI (Spain); MOSTR (Sri Lanka); Swiss Funding Agencies (Switzerland); MST (Taipei); MHESI and NSTDA (Thailand); TUBITAK and TENMAK (Turkey); NASU (Ukraine); STFC (United Kingdom); DOE and NSF (USA).
   
  \hyphenation{Rachada-pisek} Individuals have received support from the Marie-Curie programme and the European Research Council and Horizon 2020 Grant, contract Nos.\ 675440, 724704, 752730, 758316, 765710, 824093, 101115353,101002207, and COST Action CA16108 (European Union); the Leventis Foundation; the Alfred P.\ Sloan Foundation; the Alexander von Humboldt Foundation; the Science Committee, project no. 22rl-037 (Armenia); the Belgian Federal Science Policy Office; the Fonds pour la Formation \`a la Recherche dans l'Industrie et dans l'Agriculture (FRIA-Belgium); the Agentschap voor Innovatie door Wetenschap en Technologie (IWT-Belgium); the F.R.S.-FNRS and FWO (Belgium) under the ``Excellence of Science -- EOS" -- be.h project n.\ 30820817; the Beijing Municipal Science \& Technology Commission, No. Z191100007219010 and Fundamental Research Funds for the Central Universities (China); the Ministry of Education, Youth and Sports (MEYS) of the Czech Republic; the Shota Rustaveli National Science Foundation, grant FR-22-985 (Georgia); the Deutsche Forschungsgemeinschaft (DFG), under Germany's Excellence Strategy -- EXC 2121 ``Quantum Universe" -- 390833306, and under project number 400140256 - GRK2497; the Hellenic Foundation for Research and Innovation (HFRI), Project Number 2288 (Greece); the Hungarian Academy of Sciences, the New National Excellence Program - \'UNKP, the NKFIH research grants K 131991, K 133046, K 138136, K 143460, K 143477, K 146913, K 146914, K 147048, 2020-2.2.1-ED-2021-00181, and TKP2021-NKTA-64 (Hungary); the Council of Science and Industrial Research, India; ICSC -- National Research Centre for High Performance Computing, Big Data and Quantum Computing, funded by the EU NexGeneration program (Italy); the Latvian Council of Science; the Ministry of Education and Science, project no. 2022/WK/14, and the National Science Center, contracts Opus 2021/41/B/ST2/01369 and 2021/43/B/ST2/01552 (Poland); the Funda\c{c}\~ao para a Ci\^encia e a Tecnologia, grant CEECIND/01334/2018 (Portugal); the National Priorities Research Program by Qatar National Research Fund; MCIN/AEI/10.13039/501100011033, ERDF ``a way of making Europe", and the Programa Estatal de Fomento de la Investigaci{\'o}n Cient{\'i}fica y T{\'e}cnica de Excelencia Mar\'{\i}a de Maeztu, grant MDM-2017-0765 and Programa Severo Ochoa del Principado de Asturias (Spain); the Chulalongkorn Academic into Its 2nd Century Project Advancement Project, and the National Science, Research and Innovation Fund via the Program Management Unit for Human Resources \& Institutional Development, Research and Innovation, grant B37G660013 (Thailand); the Kavli Foundation; the Nvidia Corporation; the SuperMicro Corporation; the Welch Foundation, contract C-1845; and the Weston Havens Foundation (USA).  
\end{acknowledgments}

\bibliography{auto_generated}

\cleardoublepage \appendix\section{The CMS Collaboration \label{app:collab}}\begin{sloppypar}\hyphenpenalty=5000\widowpenalty=500\clubpenalty=5000
\cmsinstitute{Yerevan Physics Institute, Yerevan, Armenia}
{\tolerance=6000
A.~Hayrapetyan, A.~Tumasyan\cmsAuthorMark{1}\cmsorcid{0009-0000-0684-6742}
\par}
\cmsinstitute{Institut f\"{u}r Hochenergiephysik, Vienna, Austria}
{\tolerance=6000
W.~Adam\cmsorcid{0000-0001-9099-4341}, J.W.~Andrejkovic, T.~Bergauer\cmsorcid{0000-0002-5786-0293}, S.~Chatterjee\cmsorcid{0000-0003-2660-0349}, K.~Damanakis\cmsorcid{0000-0001-5389-2872}, M.~Dragicevic\cmsorcid{0000-0003-1967-6783}, P.S.~Hussain\cmsorcid{0000-0002-4825-5278}, M.~Jeitler\cmsAuthorMark{2}\cmsorcid{0000-0002-5141-9560}, N.~Krammer\cmsorcid{0000-0002-0548-0985}, A.~Li\cmsorcid{0000-0002-4547-116X}, D.~Liko\cmsorcid{0000-0002-3380-473X}, I.~Mikulec\cmsorcid{0000-0003-0385-2746}, J.~Schieck\cmsAuthorMark{2}\cmsorcid{0000-0002-1058-8093}, R.~Sch\"{o}fbeck\cmsorcid{0000-0002-2332-8784}, D.~Schwarz\cmsorcid{0000-0002-3821-7331}, M.~Sonawane\cmsorcid{0000-0003-0510-7010}, S.~Templ\cmsorcid{0000-0003-3137-5692}, W.~Waltenberger\cmsorcid{0000-0002-6215-7228}, C.-E.~Wulz\cmsAuthorMark{2}\cmsorcid{0000-0001-9226-5812}
% ...
\par}
\vskip\cmsinstskip
\dag:~Deceased\\
$^{1}$Also at Yerevan State University, Yerevan, Armenia\\
$^{2}$Also at TU Wien, Vienna, Austria\\
% ...
\end{sloppypar}
%%% END EDITABLE REGION %%%
% skeleton_end
\end{document}